\documentclass[sigconf]{acmart}
\setcopyright{rightsretained} 
\acmConference[Technical Report @ Arxiv]{Technical Report}{Submitted: 12. Nov. 2019}{Melbourne Australia}

\usepackage{booktabs}   %% For formal tables:
\usepackage{subcaption} %% For complex figures with subfigures/subcaptions
\usepackage{amsmath,amssymb,amsfonts,amsthm,stmaryrd}
\usepackage{algorithm,algorithmic,listings}
\usepackage{xspace}
\usepackage{graphicx}
\usepackage{textcomp}
\usepackage{xcolor}
\usepackage{multirow}
\usepackage{enumitem}
\usepackage{setspace}

\usepackage{tikz,tkz-euclide,pdftexcmds,pgfplots}
\usetikzlibrary{backgrounds}
\pgfdeclarelayer{foreground}
\pgfsetlayers{background,main,foreground}
\usetkzobj{all}

\pgfplotsset{compat=1.14}

\usepackage{tcolorbox}% http://ctan.org/pkg/tcolorbox
\definecolor{mycolor}{rgb}{0.122, 0.435, 0.698}% Rule colour
\definecolor{gray1}{gray}{0.3}

\definecolor{codegreen}{rgb}{0,0.6,0}
\definecolor{codegray}{rgb}{0.5,0.5,0.5}
\definecolor{codepurple}{rgb}{0.58,0,0.82}
\definecolor{backcolour}{rgb}{0.95,0.95,0.92}
\lstdefinestyle{mystyle}{
    commentstyle=\color{codegreen},
    keywordstyle=\color{magenta},
    numberstyle=\tiny\color{codegray},
    stringstyle=\color{codepurple},
    basicstyle=\tiny\ttfamily,
    breakatwhitespace=false,
    breaklines=true,
    captionpos=b,
    keepspaces=true,
    numbers=left,
    numbersep=5pt,
    showspaces=false,
    showstringspaces=false,
    showtabs=false,
    tabsize=2,
    columns=fixed
}
\newcommand{\result}[1]{%
\begin{tcolorbox}[colframe=mycolor,boxrule=0.5pt,arc=4pt,
      left=6pt,right=6pt,top=6pt,bottom=6pt,boxsep=0pt,width=\columnwidth]%
      {#1}
\end{tcolorbox}%
}
\definecolor{darkgreen}{rgb}{0.0, 0.5, 0.0}
\definecolor{darkred}{rgb}{0.82, 0.1, 0.26}
\newcommand{\hfd}{H\"off\-ding's inequality\xspace}

\begin{document}

%% Title information
\title{Monte Carlo Program Analysis}
\title{$(\delta,\epsilon)$-Approximate Program Analysis}
\title{Approximate Program Analysis with Probabilistic Guarantees}
\title{Probably Approximately Correct Program Analysis}
\title{Scale-Oblivious Program Analysis as PAC Learning}
\title{Scalable Greybox Program Analysis With Bounded Error}

%\title{Monte Carlo Program Analysis}
%\subtitle{Scalable Greybox Analysis With Bounded Error}
%\subtitle{Large-Scale Greybox Program Analysis With Bounded Error}

\title{MCPA: Program Analysis as Machine Learning}

%% Author information
%% Contents and number of authors suppressed with 'anonymous'.
%% Each author should be introduced by \author, followed by
%% \authornote (optional), \orcid (optional), \affiliation, and
%% \email.
%% An author may have multiple affiliations and/or emails; repeat the
%% appropriate command.
%% Many elements are not rendered, but should be provided for metadata
%% extraction tools.

%% Author with single affiliation.
\author{Marcel B{\"o}hme}
                                        %% can be repeated if necessary
\orcid{0000-0002-4470-1824}             %% \orcid is optional
\affiliation{
  \institution{Monash University, \url{marcel.boehme@acm.org}}           %% \institution is required
}
%\email{marcel.boehme@acm.org}          %% \email is recommended

%% Abstract
%% Note: \begin{abstract}...\end{abstract} environment must come
%% before \maketitle command
\begin{abstract}
%Monte Carlo methods are well-established in the sciences, such as particle physics, fluid dynamics, physical chemistry, and computational biology to solve extremely large and complex systems of equations that could never be solved with analytical, deterministic methods within a reasonable time budget.

%We argue that s
Static program analysis today takes an analytical approach which is quite suitable for a well-scoped system. Data- and control-flow is taken into account. Special cases such as pointers, procedures, and undefined behavior must be handled. A program is analyzed precisely on the statement level. 
However, the analytical approach is ill-equiped to handle implementations of complex, large-scale, heterogeneous software systems we see in the real world. Existing static analysis techniques that scale, trade correctness (i.e., soundness or completeness) for scalability and build on strong assumptions (e.g., language-specificity). Scalable static analysis are well-known to report errors that do \emph{not} exist (false positives) or fail to report errors that \emph{do} exist (false negatives). Then, how do we know the \emph{degree to which} the analysis outcome is correct?
%``how much incorrect'' the analysis outcome is?

In this paper, we propose an approach to \emph{scale-oblivious greybox program analysis with bounded error} which applies efficient approximation schemes (FPRAS) from the foundations of machine learning: PAC learnability. Given two parameters $\delta$ and $\epsilon$, with probability at least $(1-\delta)$, our \emph{Monte Carlo Program Analysis} (MCPA) approach produces an outcome that has an average error at most $\epsilon$. The parameters $\delta>0$ and $\epsilon>0$ can be choosen arbitrarily close to zero (0) such that the program analysis outcome is said to be \emph{probably-approximately correct} (PAC). %Specifically, we investigate $(\delta,\epsilon)$-approxi\-mations of the probability that one or more properties hold for an arbitrary execution of a terminating program.
We demonstrate the pertinent concepts of MCPA using three applications: $(\epsilon,\delta)$-approximate quantitative analysis, $(\epsilon,\delta)$-approximate software verification, and $(\epsilon,\delta)$-approximate patch verification.
\end{abstract}

%% 2012 ACM Computing Classification System (CSS) concepts
%% Generate at 'http://dl.acm.org/ccs/ccs.cfm'.
\begin{CCSXML}
<ccs2012>
<concept>
<concept_id>10011007.10011006.10011008</concept_id>
<concept_desc>Software and its engineering~General programming languages</concept_desc>
<concept_significance>500</concept_significance>
</concept>
<concept>
<concept_id>10003456.10003457.10003521.10003525</concept_id>
<concept_desc>Social and professional topics~History of programming languages</concept_desc>
<concept_significance>300</concept_significance>
</concept>
</ccs2012>
\end{CCSXML}

\ccsdesc[500]{Software and its engineering~General programming languages}
\ccsdesc[300]{Social and professional topics~History of programming languages}
%% End of generated code

%% Keywords
%% comma separated list
%\keywords{PAC-learnability, scale-oblivious, probabilistic, blackbox program analysis}  %% \keywords are mandatory in final camera-ready submission

%% \maketitle
%% Note: \maketitle command must come after title commands, author
%% commands, abstract environment, Computing Classification System
%% environment and commands, and keywords command.
\maketitle

\newcommand{\IEEEraisesectionheading}[1]{#1}
\newcommand{\IEEEPARstart}[2]{#1#2}
%\vspace{-0.4cm}
\IEEEraisesectionheading{\section{Introduction}}
\IEEEPARstart{I}{n}
2018, Harman and O'Hearn launched an exciting new research agenda: the innovation of frictionless\footnote{Friction is a technique-specific resistance to adoption, such as developers' reluctance to adopt tools with high false positive rates.} program analysis techniques that thrive on \emph{industrial-scale software systems} \cite{open1}. Much progress has been made. Tools like Sapienz, Infer, and Error Prone are routinely used at Facebook and Google \cite{open2,infer,errorprone}.  
Yet, \emph{many challenges remain}. For instance, the developers of Infer set the clear expectation that the tool may report many false alarms, does not handle certain language features, and can only report certain types of bugs.\footnote{\url{https://fbinfer.com/docs/limitations.html}} Such tools often trade soundness or completeness for scalability. %Special cases such as pointers, reflection, or JNI implementations, are ignored.
Then, just \emph{how sound or complete} is an analysis which ignores, e.g., expensive pointer analysis,  reflection in Java, undefined behaviors in C, or third-party libraries for which code is unavailable? %Of course, certain allowances must be made in the trade for scalability. We need to forgo expensive pointer anaysis, only focus on one procedure at a time, or ignore special cases such as Java reflection. Yet, what are the guarantees? 

% https://www.scratchapixel.com/lessons/mathematics-physics-for-computer-graphics/monte-carlo-methods-in-practice/monte-carlo-integration
We believe that static program analysis today resembles the \emph{analytical approach} in the natural sciences. However, in the natural sciences---when analytical solutions are no longer tractable---other approaches are used. \emph{Monte Carlo methods} are often the only means to solve very complex systems of equations \emph{ab initio} \cite{abinitio}. 
For instance, in quantum mechanics the following multi-dimensional integral gives the evolution of an atom that is driven by laser light, undergoes ``quantum jumps'' in a time interval $(0,t)$, and emits exactly $n$ photons at times~$t_n\ge \ldots \ge t_1$:\vspace{-0.3cm}

{\small $$
\int_0^tdt_n\int_0^{t_n}dt_{n-1}\ldots\int_0^{t_2}dt_1p_{[0,t)}(t_1,\ldots,t_n)
$$}%
where $p_{[0,t)}$ is the elementary probability density function \cite{mcappl}. 

Solving this equation is tractable only using Monte Carlo (MC) integration \cite{cannot}. MC integration solves an integral $F=\int_0^{t_i}d t_i f(\mathbf{x})$ by ``executing'' $f(\mathbf{x})$ on random inputs $\mathbf{x}\in \mathbb{R}^n$. Let $X_j\in [0,t_i]$ be the $j$-th of $N$ samples from the uniform distribution, then $F$ is estimated as $\langle F^N\rangle=\frac{t_i}{N}\sum_{j=1}^{N}f(X_j)$. MC integration \emph{guarantees} that this estimate converges to the true value $F$ at a rate of $1/\sqrt{N}$. This is true, no matter how many variables $t_i$ the integral has or how the function $f:\mathbb{R}^n\rightarrow \mathbb{R}$ is ``implemented''.

\begin{figure}[t]\footnotesize
\begin{tabular}{@{}l@{ \ }r@{ }l@{ \ }l@{ }l@{}|r@{ \ \ }l@{}}
 \multicolumn{5}{@{}c|}{\textbf{User-Generated Executions Per Day}}& \multicolumn{2}{c}{\textbf{Accuracy} $\mathbf{\epsilon}$}\\
 & & & & & \multicolumn{1}{c}{HFD} & \multicolumn{1}{c}{Ro3}\\\hline
OSS-Fuzz &  200.0 & \emph{billion} & test inputs* &\cite{oss} & $3.6\cdot 10^{-6}$ & $2.3\cdot 10^{-11}$\\
Netflix  &   86.4 & \emph{billion} & API requests &\cite{netflixStat} &$5.5\cdot 10^{-6}$ & $5.3\cdot 10^{-11}$\\
Youtube  &    6.2 & \emph{billion} & videos watched &\cite{livestats2}&$2.1\cdot 10^{-5}$ & $7.4\cdot 10^{-10}$\\
Google   &    5.6 & \emph{billion} & searches &\cite{livestats2}&$2.1\cdot 10^{-5}$ & $8.2\cdot 10^{-10}$\\
Tinder   &    2.0 & \emph{billion} & swipes &\cite{tinder} &$3.6\cdot 10^{-5}$& $2.3\cdot 10^{-9}$\\
%Snapchat &    3.0 & \emph{billion} & new snaps \cite{livestats2}&$3.9\cdot 10^{-5}$& $1.5\cdot 10^{-9}$\\
Facebook &    1.5 & \emph{billion} & user logins &\cite{fbStat}&$4.2\cdot 10^{-5}$& $3.1\cdot 10^{-9}$\\
Twitter  &  681.7 & \emph{million} & tweets posted &\cite{livestats2}&$6.2\cdot 10^{-5}$& $6.8\cdot 10^{-9}$\\
Skype    &  253.8 & \emph{million} & calls &\cite{livestats2}&$1.0\cdot 10^{-4}$& $1.8\cdot 10^{-8}$\\
Visa     &  150.0 & \emph{million} & transactions &\cite{visa} & $1.3\cdot 10^{-4}$ & $3.1\cdot 10^{-8}$\\
Instagram &  71.1 & \emph{million} & photos uploaded &\cite{livestats2}&$1.9\cdot 10^{-4}$&$6.5\cdot 10^{-8}$\\
%Reddit   &    2.8 & \emph{million} & new comments &\cite{livestats2}& $9.7\cdot 10^{-4}$&$1.6\cdot 10^{-6}$\\
%Uber     &    2.0 & \emph{million} & rides booked \cite{livestats2}& $1.2\cdot 10^{-3}$&$2.3\cdot 10^{-6}$\\\hline
\multicolumn{7}{@{}l@{}}{(*) per widely-used, security-critical library, on average. 10 \emph{trillion} total.}\\[-0.27cm]
\end{tabular}
\caption{Daily number of executions of large software systems in 2018, and the lower (Ro3) and upper bounds (HFD) on the accuracy $\epsilon$ of an estimate $\hat\mu$ of the probability $\mu$ that a binary program property $\varphi$ holds (e.g.,  bug is found). Specifically, $\mu\in [\hat\mu-\epsilon, \hat\mu+\epsilon]$ with probability $\delta=0.01$ (i.e., 99\%-CIs).\vspace{-0.32cm}}
\label{fig:realworld}
\end{figure}

We argue that frictionless program analysis at the large, industrial scale requires a fundamental change of perspective. Rather than starting with a precise program analysis, and attempting to carefully trade some of this precision for scale, we advocate an inherently \emph{scale-oblivious} approach. 
In this paper, we cast scale-oblivious program analysis as a probably-approximately correct (PAC)-learning problem \cite{paclearning}, provide the probabilistic framework, and develop several fully polynomial-time randomized approximation schemes (FPRAS). Valiant \cite{valiant84} introduced the PAC framework to study the computational complexity of machine learning techniques. The objective of the learner is to receive a set of samples and generate a hypothesis that with high probability has a low generalization error. We say the hypothesis is probably, approximately correct, where both adverbs are formalized and quantified. 

We call our technique \emph{Monte Carlo program analysis} (MCPA). The \emph{learner} is the program analysis. The \emph{samples} (more formally Monte Carlo trials) are distinct program executions that are potentially generated by different users of the software system under normal workload. The \emph{hypothesis} is the analysis outcome. Given two parameters $\delta$ and $\epsilon$, an MCPA produces an outcome that with probability at least $(1-\delta)$ has an average error at most $\epsilon$. It is important to note that the analysis outcome is probably-approximately correct w.r.t. the distribution of Monte Carlo trials. For instance, if the executions are generated by real users, the analysis outcome is probably-approximately correct w.r.t. the software system as it is used by real users.

In contrast to existing program analysis techniques, MCPA\vspace{-0.1cm}
\begin{itemize}[itemsep=1pt]
  \item allows to reason about the entire software systems \emph{under normal workload}, i.e., an MCPA \emph{while} the system is used normally produces an outcome w.r.t. normal system use,
  \item requires only \emph{greybox access} to the software system, i.e., no source code is analyzed, instead lightweight program instrumentation allows to monitor software properties of interest,
  \item provides \emph{probabilistic guarantees} on the correctness and accuracy of the analysis outcome; i.e., to exactly quantify the scalability-correctness tradeoff,
  \item can be \emph{implemented in a few lines} of Python code, and
  \item is massively parallelizable, scale-oblivious, and can be interupted at any time.
\end{itemize}
%\newpage

MCPA is a general approach that can produce outcomes with bounded error for arbitrary program analyses. This is because the PAC-framework covers all of machine learning which can learn arbitrary concepts, as well. Shalev-Shwartz and Ben-David \cite{mlshalev} provide an excellent overview of the theory of machine learning.

However, in this paper we focus here on $(\epsilon,\delta)$-approxi\-mations of the probability that a \emph{binary property holds} for an arbitrary execution of a terminating program. This has important applications, e.g., in assessing reliability \cite{filieri1,filieri2,filieri3}, quantifying information leaks \cite{qif}, or exposing side-channels \cite{bultan}. Given parameters $\delta>0$ and $\epsilon>0$, our MCPA \emph{guarantees} that the reported estimate $\hat\mu$ is no more than $\epsilon$ away from the true value $\mu$ with probability at least $1-\delta$, i.e., $P(|\hat \mu - \mu|\ge \epsilon)\le \delta$.  In statistical terms, MCPA \emph{guarantees} the $(1-\delta)$-confidence interval $\hat \mu\pm\epsilon$. Note that confidence and accuracy parameters $\delta$ and $\epsilon$ can be chosen \emph{arbitrarily close to zero} (0) to minimize the probability of false negatives.%\footnote{Here we should note that this specific objective is shared with probabilistic and statistical model checking. However, unlike such runtime verification, MCPA is neither focused on temporal properties nor concerned with a program's state space. More specifically, instead of verifying the expanding prefix of an infinite path through the program's state space, MCPA requires several distinct, terminating executions (Sec.~\ref{sec:related}). In this setting, the work on probabilistic symbolic execution \cite{pse,filieri1,filieri2} is most related.}

To illustrate the power of MCPA, we show how often well-known, industrial-scale software systems are executed \emph{per day} (Fig.~\ref{fig:realworld}). For instance, Netflix handles 86.4 \emph{billion} API requests ($8.64\cdot 10^{10}$) per day.
MCPA \emph{guarantees} that the following statements are true with probability greater than 99\% w.r.t. the sampled distribution:
\begin{itemize}[itemsep=2pt]
  \item No matter which binary property we check or how many properties we check simultaneously, we can guarantee \emph{a priori} that the error $\epsilon$ of our estimate(s) $\hat\mu$ of $\mu$ is bounded between $5.3\cdot 10^{-11} \le \epsilon \le 5.3\cdot 10^{-6}$ such that $\mu\in\hat\mu\pm\epsilon$.
  \item If 0 out of 86.4 \emph{billion} API requests expose a security flaw, then the \emph{true} probability that there exists a security flaw that has not been exposed is less than $5.3\cdot 10^{-11}$ (Sec.~\ref{sec:verif}).
  	\item If 1000 of 86.4 \emph{billion} API requests expose a vulnerability, then the \emph{true} probability $\mu$ is probabilistically guaranteed to be within $\mu\in 1.16\times 10^{-8} \pm 1.45\times 10^{-9}$ (Sec. \ref{sec:quantbin}).
  \item If a patch was submitted, it would require \emph{at most} 400 \emph{million} API requests (equiv. 7 minutes) to state with confidence (\emph{p-value}$<10^{-3}$) that the failure probability has indeed reduced if none of those 400M executions exposes the bug \emph{after} the patch is applied. MCPA provides an \emph{a-priori conditional probabilistic guarantee} that the one-sided null-hypothesis can be rejected at significance level $\alpha=0.01$ (Sec. \ref{sec:patchverif}).
  \item For an arbitrary program analysis, let $\mathcal{H}$ be a finite hypothesis class containing all possible program analysis outcomes. If MCPA chooses the outcome $h\in \mathcal{H}$ which explains the analysis target (e.g., points-to-analysis) for all 5.6 \emph{billion} daily Google searches, then $h$ explains the analysis targets with a maximum error of $\epsilon\le\frac{\log(|H|/\delta)}{5.6\cdot 10^9}$ for ``unseen'' searches \cite{mlshalev}.
  
\end{itemize}

On a high level, MCPA combines the advantages of static and dynamic program analysis while mitigating  individual challenges.
Like a static program analysis, MCPA allows to make statements over all executions of a program. However, MCPA does \emph{not} take an analytical approach. A static analysis evaluates statements in the program source code which requires strong assumptions about the programming language and its features. This poses substantial challenges for static program analysis. For instance, alias analysis is undecidable \cite{alias}. 
Like a dynamic program analysis, MCPA overcomes these challenges by analyzing the \emph{execution} of the software system. However, MCPA does not analyze only one execution. Instead, MCPA allows to make statements over all executions generated from a given (operational) distribution.

\vspace{0.1cm}
\noindent
The \emph{contributions} of this article are as follows:\vspace{-0.1cm}
\begin{itemize}[leftmargin=10pt]
  \item We discuss the opportunities of employing Monte Carlo methods as an approximate, greybox approach to large-scale quantitative program analysis with bounded error.  
  \item We develop several fully-polynomial randomized approximation schemes (FPRAS) that guarantee that the produced estimate $\hat\mu$ of the probability $\mu$ that a program property holds is \emph{no more than} $\epsilon$ away from $\mu$ with probability \emph{at least} $(1-\delta)$. We tackle an open problem in automated program repair: When is it appropriate to claim with some certainty that the bug is indeed repaired? 
  \item We evaluate the efficiency of our MCPA algorithms probabilistically by providing upper and lower bounds \emph{for all} $0\le\mu\le 1$, and empirically in more than $10^{18}$ simulation experiments.%\item A survey of fully-polynomial randomized approximation schemes (FPRAS) from the complexity-theoretic foundations of machine learning: probably approximately correct (PAC) learnability.
\end{itemize}
%\newpage

%\input{sections/background}
\section{Monte Carlo Program Analysis}\label{sec:mcpa}
Monte Carlo program analysis (MCPA) allows to derive statements about properties of a software system that are probably approximately correct \cite{valiant}. Given a confidence parameter $0<\delta <1$ and an accuracy parameter $\epsilon>0$, with probability at least $(1-\delta)$, MCPA produces an analysis outcome that has an average error at most $\epsilon$.

\textbf{Problem statement}. In the remainder of this paper, we focus on a special case of MCPA, i.e., to assess the \emph{probability} $\mu$ that a \emph{binary property} $\varphi$ holds (for an arbitrary number of such properties). The probability that $\varphi$ does \emph{not} hold (i.e., that $\neg\varphi$ holds) is simply $(1-\mu)$.
Suppose, during the analysis phase, it was observed that $\varphi$ holds for the proportion $\hat\mu$ of sample executions. We call those sample executions in the analysis phase as \emph{Monte Carlo trials}. The analysis phase can be conducted during testing or during normal usage of the software system. The analysis outcome of an MCPA is expected to hold w.r.t. the \emph{concerned executions}, e.g., further executions of the system during normal usage. 
Our objective is to guarantee for the concerned executions that the reported estimate $\hat\mu$ is no more than $\epsilon$ away from the true value $\mu$ with probability at least $1-\delta$, i.e., $P(|\hat \mu - \mu|\ge \epsilon)\le \delta$. 

\textbf{Assumptions}.
We adopt the classical assumptions from software testing: We assume that (i)~all executions \emph{terminate}, (ii)~property violation is \emph{observable} (e.g., an uncaught exception), and (iii)~the Monte Carlo trials are representative of the concerned executions. We elaborate each assumption and their consequences in the following sections.
We make no other assumptions about the program $\mathcal{P}$ or the property $\varphi$. For instance, $\mathcal{P}$ could be deterministic, probabilistic, distributed, or concurrent, written in C, Java, or Haskell, or a trained neural network, or a plain old program. We could define $\varphi$ to hold if an execution yields correct behavior, if a security property is satisfied, if a soft deadline is met, if an energy threshold is not exceeded, if the number of cache-misses is sufficiently low, if no buffer overwrite occurs, if an assertion is not violated, et cetera.

\vspace{-0.2cm}
\subsection{Assumption 1: All Executions Terminate}
We assume that all executions of the analyzed program $\mathcal{P}$ \emph{terminate}. With terminating executions we mean independent units of behavior that have a beginning and an end. For instance, a user session runs from login to logout, a transaction runs from the initial handshake until the transaction is concluded, and a program method runs from the initial method call until the call is returned. Other examples are shown in \autoref{fig:realworld}. This is a realistic assumption also in software testing.

\textbf{If assumption does not hold}.
As we require all (sample and concerned) executions to terminate, MCPA cannot verify temporal properties, such as those expressed in linear temporal logic (LTL). If the reader wishes to verify temporal properties over non-terminating executions, we refer to probabilistic model checking \cite{pmc,probmodelcheck1,probmodelcheck2} or statistical model checking \cite{smc,sen}. If the reader wishes to verify binary properties over non-terminating executions, we suggest to use probabilistic symbolic execution \cite{pse} on partial path constraints of bounded length $k$.

\vspace{-0.2cm}
\subsection{Assumption 2: Property Outcomes Are Observable}
We assume that the outcome of property $\varphi\in \{0,1\}$ can be automatically observed for each execution.
Simply speaking, we cannot estimate the proportion $\hat\mu$ of concerned executions for which $\varphi$ holds, if we cannot observe whether $\varphi$ holds for any Monte Carlo trial during the analysis phase. This is a realistic assumption that also exists in software testing.

\textbf{Greybox access}. Some properties $\varphi$ can be observed externally without additional code instrumentation. For instance, we can measure whether latency, performance, or energy thresholds are exceeded by measuring the time it takes to respond or to compute the final result, or how much energy is consumed.
Other properties can be observed by injecting lightweight instrumention directly into the program binary, e.g., using DynamoRIO or Intel Pin. Other properties can be made observable at compile-time causing very low runtime overhead. An example is AddressSanitizer \cite{asan} which is routinely compiled into security-critical program binaries to report (exploitable) memory-related errors, such as buffer overflows and use-after-frees. 

\vspace{-0.2cm}
\subsection{Assumption 3: Property Outcomes Are Independent and Identically Distributed}
The sequence of $n$ MC trials is a stochastic process $\mathcal{F}\{X_m\}_{m=1}^n$ where $X_m\in \mathcal{F}$ is a binomial random variable that is true if $\varphi=1$ in the $m$-th trial. We assume that the property outcomes $\mathcal{F}$ are independent and identically distributed (\emph{IID}). This is a classic assumption in testing, e.g., all test inputs are sampled \emph{independently} from the \emph{same} (operational) distribution \cite{filieri1}. More generally, throughout the analysis phase, the probability $\mu$ that $\varphi$ holds is invariant; the binomial distribution over $X_m$ is the same for all $X_m\in \mathcal{F}$.

\textbf{Testing IID}. In order to test whether $\mathcal{F}$ is IID, there are several statistical tools available. For instance, the \emph{turning point test} is a statistical test of the independence of a series of random variables. %It can be used to test the null hypothesis that a sequence of property outcomes are IID.

\textbf{If assumption does not hold}. The assumption that executions are identically distributed does not hold for stateful programs where the outcome of $\varphi$ in one execution may depend on previous executions. In such cases, we suggest to understand each fine-grained execution as a transition from one state to another within a single non-terminating execution. The state transitions can then be modelled as Markov chain and checked using tools such as probabilistic model checking \cite{probmodelcheck1,probmodelcheck2}.

\vspace{-0.2cm}
\subsection{Assumption 4: Monte Carlo Trials Represent Concerned Executions}
We assume that the Monte Carlo trials are representative of the concerned executions. In other words, the executions of the software system that were generated during the analysis phase are from the same distribution as the executions w.r.t. which the analysis outcome is expected to hold. This is a realistic assumption shared with software testing, and any empirical analysis in general.

\textbf{Realistic Behavior}. A particular strength of MCPA compared to existing techniques is that the software system can be analyzed under normal workload to derive an probably-approximately correct analysis outcome that holds w.r.t. the software system as it is normally executed.

\section{Motivation: Challenges of the Analytical Approach}\label{sec:motivation}
The state-of-the-art enabling technology for quantitative program analysis is \emph{probabilistic symbolic execution} (PSE) which combines symbolic execution and model counting.  In a 2017 LPAR keynote \cite{lpar}, Visser called PSE the \emph{new hammer}. Indeed, we find that PSE is an exciting, new tool in the developer's reportoire for quantitative program analysis problems particularly if an exact probability is required. However, the analytical approach of PSE introduces several challenges for the analysis of large-scale, heterogeneous software systems which MCPA is able to overcome.

\subsection{Probabilistic Symbolic Execution}
Suppose property $\varphi$ is satisfied for all program paths $I$. Conceptually, for each path $i\in I$, PSE computes the probability $p_i$ that an input exercises $i$ and then reports $\mu=\sum_{i\in I}p_i$. 
To compute the probability $p_i$ that an input exercises path $i$, PSE first uses \emph{symbolic execution} to translate the source code that is executed along $i$ into a Satisfiability Modulo Theory (SMT) formula, called path condition. The \emph{path condition} $\pi(i)$ is satisfied by all inputs that exercise $i$. A \emph{model counting} tool can then determine the number of inputs $\llbracket \pi(i)\rrbracket$ that exercise $i$. Given the size $\llbracket D\rrbracket$ of the entire input domain $D$, the probability $p_i$ to exercise path $i$ can be computed as  $p_i=\llbracket \pi(i)\rrbracket/\llbracket D\rrbracket$.

There are two approaches to establish the model count. Exact model counting \cite{pse,filieri1,bultan,bultan2} determines the model count \emph{precisely}. For instance, LaTTE \cite{latte} computes $\llbracket \pi(i)\rrbracket$ as the volume of a convex polytope which represents the path constraint. However, an exact count \emph{cannot} be established efficiently and turns out to be intractable in practice \cite{sharpp,sharpSMT}.\footnote{Gulwani et al. \cite{gulwani} observe that the ``%We observed two limitations of LattE that make it less ideal for probabilistic program analysis tasks: (a) the existing implementation does not handle non-uniform distributions (over the reals or integers), and (b) 
exact volume determination for real polyhedra is quite expensive, and often runs out of time/memory on the constraints that are obtained from our benchmarks''. %They ``attempted to complete at least one of the benchmark examples in our approach using LattE, but were unsuccessful due to timeouts or more often out-of-memory errors''. 
%Moreover, t
Time is exponential in the number of input variables. While a simple constraint involving four variables is quantified in less then 3 \emph{minutes} in some experiments, a similarly simple constraint involving eight variables is quantified in just under 24 \emph{hours} \cite{latte}.} Hence, recent research has focussed on PSE involving approximate model counting.

Approximate model counting \emph{trades accuracy for efficiency} and determines $\llbracket \pi(i)\rrbracket$ approximately. Inputs are sampled from non-overlapping, axis-aligned bounding boxes that together contain all (but not only) solutions of $\pi(i)$ \cite{filieri2,filieri3,filieri4,filieri5}. To determine whether the sampled input exercises $i$, it is plugged into $\pi(i)$ and checked for satisfiability. 
The proportion of ``hits'' multiplied by the size of the bounding box, summed over all boxes gives the approximate model count. In contrast to arbitrary polytopes, the size of an axis-aligned bounding box can be precisely determined efficiently.

\subsection{Challenges}
Probabilistic symbolic execution uses an incremental encoding of the program's source code as a set of path conditions. Whether static or dynamic symbolic execution, the semantic meaning of each individual program statement is translated into an equivalent statement in the supported Satisfiability Modulo Theory (SMT). The quantitative program analysis, whether exact or approximate, is then conducted upon those path conditions.
This indirection introduces several limitations.
\begin{enumerate}[itemsep=2pt]
  \item Only programs that can be represented in the available SMT theory can be analyzed precisely (e.g., bounded linear integer arithmetic \cite{pse} or strings \cite{bultan2}). For approximate PSE, only bounded numerical input domains are allowed. Otherwise,  bounding boxes and their relative size \emph{cannot} be derived. Programs that are built on 3rd-party libraries, are implemented in several languages, contain unbounded loops, involve complex data structures, execute several threads simultaneously pose challenges for probabilistic symbolic execution.
  \item Only functional properties $\varphi$ can be checked; those can be encoded within a path condition, e.g., as assertions over variable values. In contrast, non-functional properties, such as whether the response-time or energy consumption exceeds some threshold value, are difficult to check.
  \item Only a lower bound on the probability $\mu$ that $\varphi$ holds can be provided. PSE cannot efficiently enumerate \emph{all} paths for which property $\varphi$ holds. Reachability is a hard problem. Otherwise, symbolic execution would be routinely used to prove programs correct. Hence, PSE computes $\mu$ as $\mu' = \sum_{i\in I'}p_i$ only for a reasonable subset of paths $I'\subseteq I$ that satisfy $\varphi$ \cite{filieri1}, which is why the reported probability $\mu'\le\mu$.
  \item No confidence intervals are available for approximate PSE. While approximate PSE trades accuracy (of the analysis outcome) for efficiency (of the analysis), it does not provide any means to assess this trade-off.\footnote{We note that the sample variance $\hat\sigma$ is \emph{not} an indicator of estimator accuracy. Suppose, $\mu=0.5$ such that the (true) variance $\sigma^2=0.25/n$ is maximal. Due to randomness, we may take $n$ samples where $\varphi$ does not hold. Our estimate $\hat \mu=0$ is clearly inaccurate, yet the sample variance $\hat\sigma^2=0$ gives no indication of this inaccuracy.} Approximate PSE provides only a (negatively biased) point estimate $\hat \mu$ of $\mu$. Due to the modular nature of the composition of the estimate $\hat \mu$, it is difficult to derive a reliable $(1-\delta)$-confidence interval $[\hat\mu-\epsilon,\hat\mu+\epsilon]$. %There exits no stopping rule to determine how many samples needed from each bounding box to gain a final estimate $\hat\mu$ that is sufficiently accurate.
  \item To provide an analysis outcome for programs under normal workload, a usage profile must be developed that can be integrated with the path condition. PSE requires to formally encode user behavior as a usage profile \cite{filieri1}. However, deriving the usage profile from a sample of typical execution can introduce uncertainty in the stochastic model that must be accounted for \cite{yamilet}.
\end{enumerate}
We believe that many of these limitations can be addressed. Yet, they do demonstrate a shortcoming of the analytical approach. A static program analysis must parse the source code and interpret each program statement on its own. It cannot rely on the compiler to inject meaning into each statement. The necessary assumptions limit the applicability of the analysis. The required machinery is a substantial performance bottleneck. For PSE, we failed to find reports of experiments on programs larger than a few hundred lines of code. Notwithstanding,
we are excited about recent advances in scalable  program analysis where correctness is carefully traded for scalability. However, how do we quantify the error of the analysis outcome? What is the probability of a false positive/negative? 

\subsection{Opportunities}
Monte Carlo program analysis resolves many of these limitations.
\begin{enumerate}[itemsep=2pt]
  \item Every executable program can be analyzed. Apart from termination we make no assumptions about the software system.
  \item Every binary property $\varphi$ can be checked, including a non-functional property, as long as it can be automatically observed. Moreover, the number of required trials is independent of the number of simultaneously checked properties.
  \item The maximum likelihood estimate $\hat\mu$ for the binomial proportion $\mu$ is unbiased. Unlike PSE, MCPA does not require to identify or enumerate program paths that satisfy $\varphi$.
  \item The error is bounded. The analysis results are probably approximately correct, i.e., given the parameters $\delta$ and $\epsilon$, with probability $(1-\delta)$, we have that $\mu \in [\hat\mu-\epsilon,\hat\mu+\epsilon]$.
  \item The best model of a system is the running system itself. The system can be analyzed directly during normal execution with analysis results modulo concerned executions.
\end{enumerate}
It is interesting to note that approximate probabilistic symbolic execution employs Monte Carlo techniques upon each path condition, effectively \emph{simulating} the execution of the program. Our key observation is that the compiler already imbues each statement with meaning and program binaries can be executed concretely many orders of magnitutes faster than they can be simulated.

 \section{$(\epsilon,\delta)$-Approximate Software Verification}\label{sec:verif}

During the analysis of large software systems, the probablity $\mu$ that a binary program property $\varphi$ holds for an execution is often \emph{almost or exactly one}.\footnote{Note that we can reduce the alternative case---where the probablity $\mu$ that $\varphi$ holds is almost or exactly zero---to the case that we consider here. In that case, the probablity $(1-\mu)$ that $\varphi$ does \emph{not} hold (i.e., $\neg\varphi$ holds) is almost or exactly one.} For a very large number of successive Monte Carlo trials, we might not ever observe that $\varphi$ is violated. Still, there is always some residual risk that $\neg\varphi$ holds for an infinitesimal proportion of concerned executions, i.e., $(1-\mu)>0$ \cite{assurances}. Hence, $(\epsilon,\delta)$-\emph{approximate software verification} allows to specify an \emph{allowable residual risk} $\epsilon$. It guarantees $0\le(1-\mu)\le\epsilon$ with probability $(1-\delta)$.

If the binary program property $\varphi$ holds for all of $n$ Monte-Carlo trials, our empirical estimate $\hat \mu$ of the probability $\mu$ that $\varphi$ holds is $\hat \mu=\frac{n}{n}=1$. In this section, we provide a better point estimate $\hat\mu_L$ for $\mu$ in the absence of empirical evidence for the violation of $\varphi$. Now, \hfd already provides probabilistic error bounds (Sec. \ref{sec:quantbin}): Given the confidence parameter $\delta$, we can compute the accuracy $\epsilon$ of the estimate $\hat\mu$ such that $\mu$ is within $\hat\mu\pm\epsilon$ with probability $(1-\delta)$. However, in this special case  we can leverage a generalization of the rule-of-three (Ro3) to reduce the radius $\epsilon$ of the guaranteed confidence interval by many orders of magnitude.

\begin{algorithm}[t]
\caption{$(\epsilon,\delta)$-Approximate Software Verification}\label{alg:mcpa2}
\renewcommand{\algorithmicrequire}{\textbf{Input:}}
\renewcommand{\algorithmicensure}{\textbf{return}}
\begin{algorithmic}[1]
%\REQUIRE Program $\mathcal{P}$, Inputs $T$ \textcolor{darkgray}{(where $|T|=n$)}
%\REQUIRE Properties $\Phi$, Confidence $\delta$
\REQUIRE Program $\mathcal{P}$, Binary properties $\Phi$
\REQUIRE Accuracy $\epsilon$, Confidence $\delta$
\STATE Let $n = \log(\delta) / \log(1-\epsilon)$
\FOR{Each of $n$ Monte Carlo trials}
  \STATE Execute program $\mathcal{P}$ and observe property outcomes
  \FOR{$\varphi_i\in \Phi$}
    \STATE \textbf{if} $\varphi_i$ does \emph{not} hold in current execution
    \STATE \textbf{then return} \emph{``Property $\varphi_i$ violated for current execution $\mathcal{P}$''} 
   \ENDFOR
\ENDFOR
\ENSURE \emph{``For all $\varphi_i\in \Phi$, we \textbf{estimate} the probability that $\varphi_i$ holds in $\mathcal{P}$ is $\hat\mu_L=(n+1)/(n+2)$ and \textbf{guarantee} that the \emph{true} probability $\mu_i \in [1-\epsilon,1]$ with probability at least $(1-\delta)$.''}
%\ENSURE Estimates $\hat \mu_i$; Accuracy $\epsilon$
\end{algorithmic}
\end{algorithm}

\textbf{Point estimate}. Given that the sun has risen ever since we were born $n$ days ago, what is the probability that the sun will rise tomorrow? This riddle is known as the \emph{sunrise problem} and the solution is due to Laplace. Obviously, our plugin estimator $\hat\mu=1$ is positively biased, as there is always some residual probability that the sun will fail to rise tomorrow (i.e., that $\neg\varphi$ holds). If we have observed that $\varphi$ holds in all of $n$ trials, the \emph{Laplace estimate} $\hat\mu_L$ of $\mu$ is computed as%\vspace{-0.2cm}
\begin{align}
\hat\mu_L = \frac{n+1}{n+2}
\end{align}

\textbf{Error bounds}. The \emph{rule-of-three} \cite{hanley} is widely used in the evaluation of medical trials. It yields the 95\%-confidence interval $[0,3/n]$ for the probability of adverse effects given a trial where none of the $n$ participants experienced adverse effects.
Given an \emph{accuracy parameter} $\epsilon>0$, Hanley and Lippmann-Hand \cite{hanley} provide an upper bound on the probability that the estimate $\hat \mu$ deviates more than $\epsilon$ from the true value $\mu$, and since $\hat \mu=1$ also on the probability that $\mu\ge 1-\epsilon$.
\begin{align}
P(|\hat \mu - \mu|\ge \epsilon) \le (1-\epsilon)^n
\end{align}
Given a \emph{confidence parameter} $\delta:0<\delta\le 1$, we can compute the number $n$ of Monte Carlo trials required to guarantee the $(1-\delta)$-confidence interval $[\hat \mu-\epsilon,\hat \mu+\epsilon]$ as
\begin{align}
n \ge \frac{\log(\delta)}{\log(1-\epsilon)}\label{eq:eight}
\end{align}
Analogously, given $n$ Monte Carlo trials, with probability at least $1-\delta$, the absolute
difference between the estimate $\hat\mu$ and the true value $\mu$ is at most $\epsilon$ where
\begin{align}
\epsilon \le 1-\delta^{\frac{1}{n}}\label{eq:nine}
\end{align}
Note that for a 95\%-CI (i.e., $\delta=0.05$), we have that $\epsilon\le 3/n$ giving rise to the name rule-of-three.
 
\begin{figure}\centering
\includegraphics[width=0.52\columnwidth, clip=true,trim= 0 0 0 0.8cm]{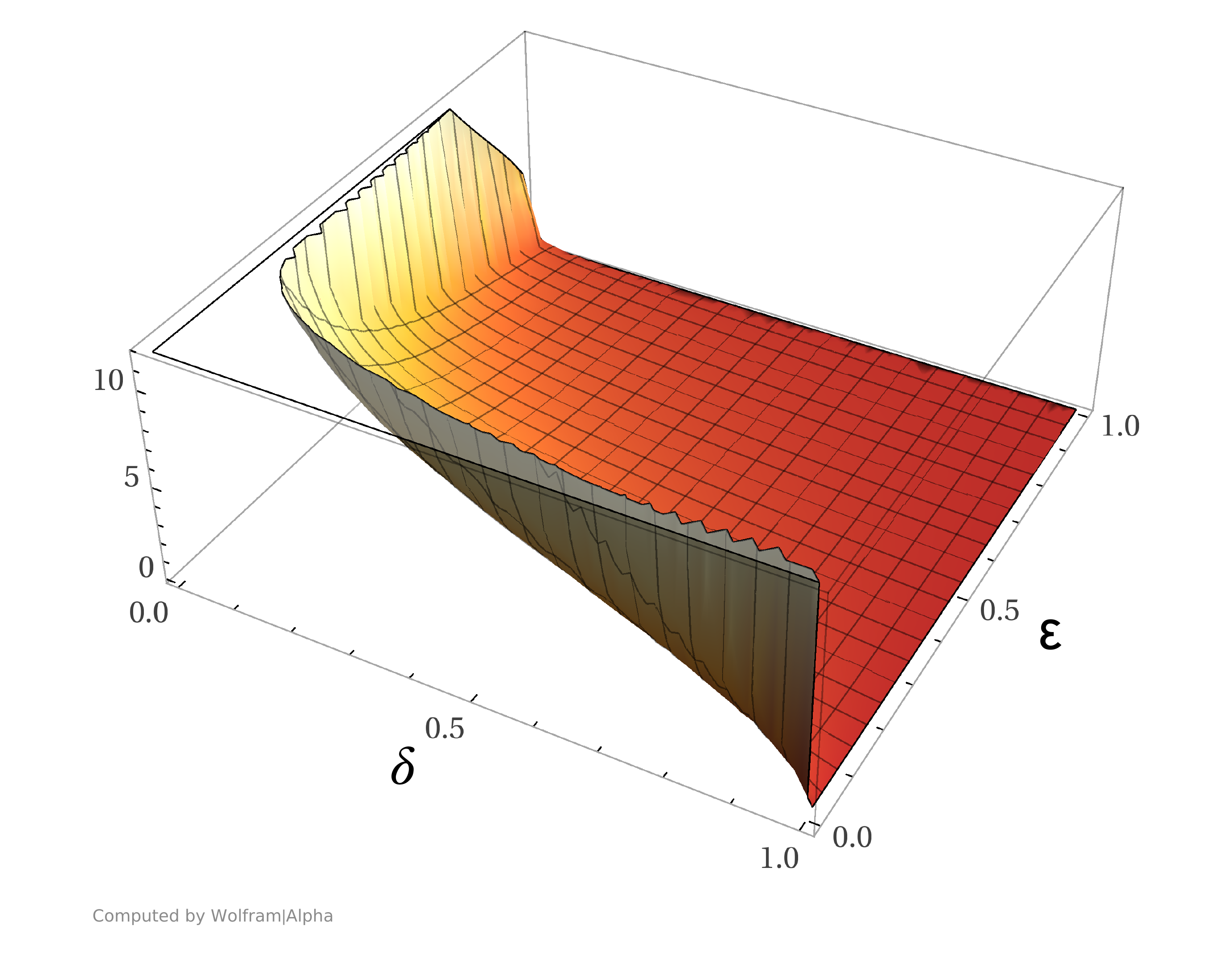}
\includegraphics[width=0.42\columnwidth]{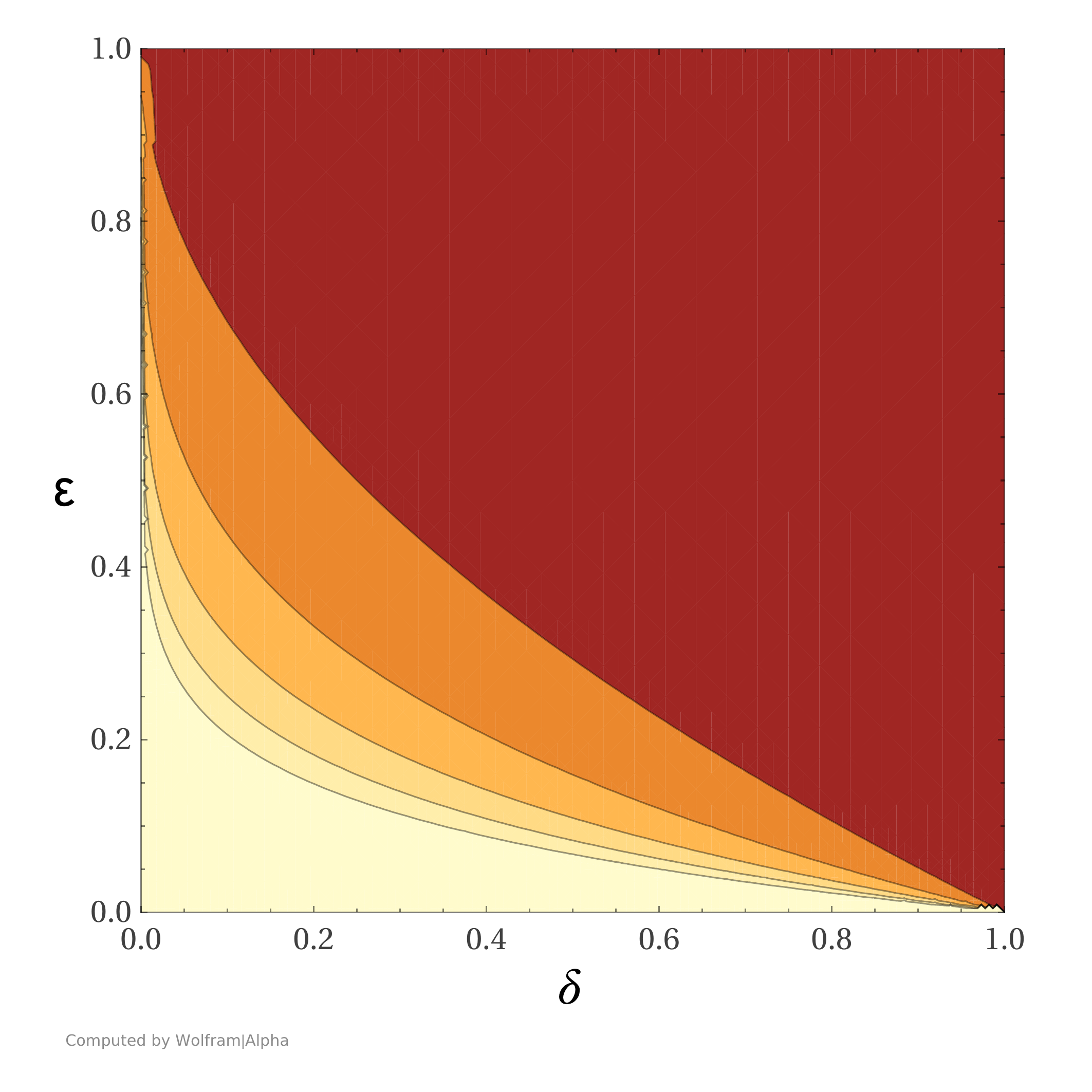}\\[-0.1cm]
\tiny\color{gray}
Wolfram Alpha LLC. 2019. \url{https://www.wolframalpha.com/input/?i=plot+log(delta)/log(1-epsilon)+for+delta=0..1,+epsilon=0..1} (access June 23, 2019).
\caption{The number $n$ of Monte Carlo trials required, as confidence $\delta$ and accuracy $\epsilon$ varies. The surface plot (left) is a 3D plot of the function. The contour plot (right) visualizes how $n$ changes as $\delta$ and $\epsilon$ varies.}
\label{fig:ro3plots}
\end{figure}

\textbf{Efficiency}.
Our Algorithm~\ref{alg:mcpa2} runs in time that is polynomial in $\log(1/(1-\epsilon)))^{-1}<1/\epsilon$ and in $\log(1/\delta)$. It is thus a fully polynomial randomized approximation scheme (FPRAS). The efficiency of Algorithm~\ref{alg:mcpa2} is visualized in \autoref{fig:ro3plots}. To reduce $\epsilon$ by an order of magnitude increases the required Monte Carlo trials also only by an order of magnitude. This is further demonstrated for the real-world examples shown in \autoref{fig:realworld}.

\section{$(\epsilon,\delta)$-Approximate Quantitative Analysis}\label{sec:quantbin}
Quantitative program analysis is concerned with the proportion $\mu$ of executions that satisfy some property $\varphi$ of interest. Quantitative analysis has important applications, e.g., in assessing software reliability \cite{filieri1,filieri2,filieri3}, computing performance distributions \cite{bihuan}, quantifying information leaks \cite{qif}, quantifying the semantic difference across two program versions \cite{changes}, or exposing side-channels \cite{bultan}. 

We propose $(\epsilon,\delta)$-\emph{approximate quantitative analysis} which yields an estimate $\hat\mu$ that with probability at least $(1-\delta)$ has an average error at most $\epsilon$. While the generalized rule of three (Ro3) provides a lower bound on the number  trials needed to compute an $(\epsilon,\delta)$-approximation of $\mu$, \hfd provides an upper bound.

\begin{figure}[t]\centering
\includegraphics[width=0.52\columnwidth, clip=true,trim= 0 0 0 0.8cm]{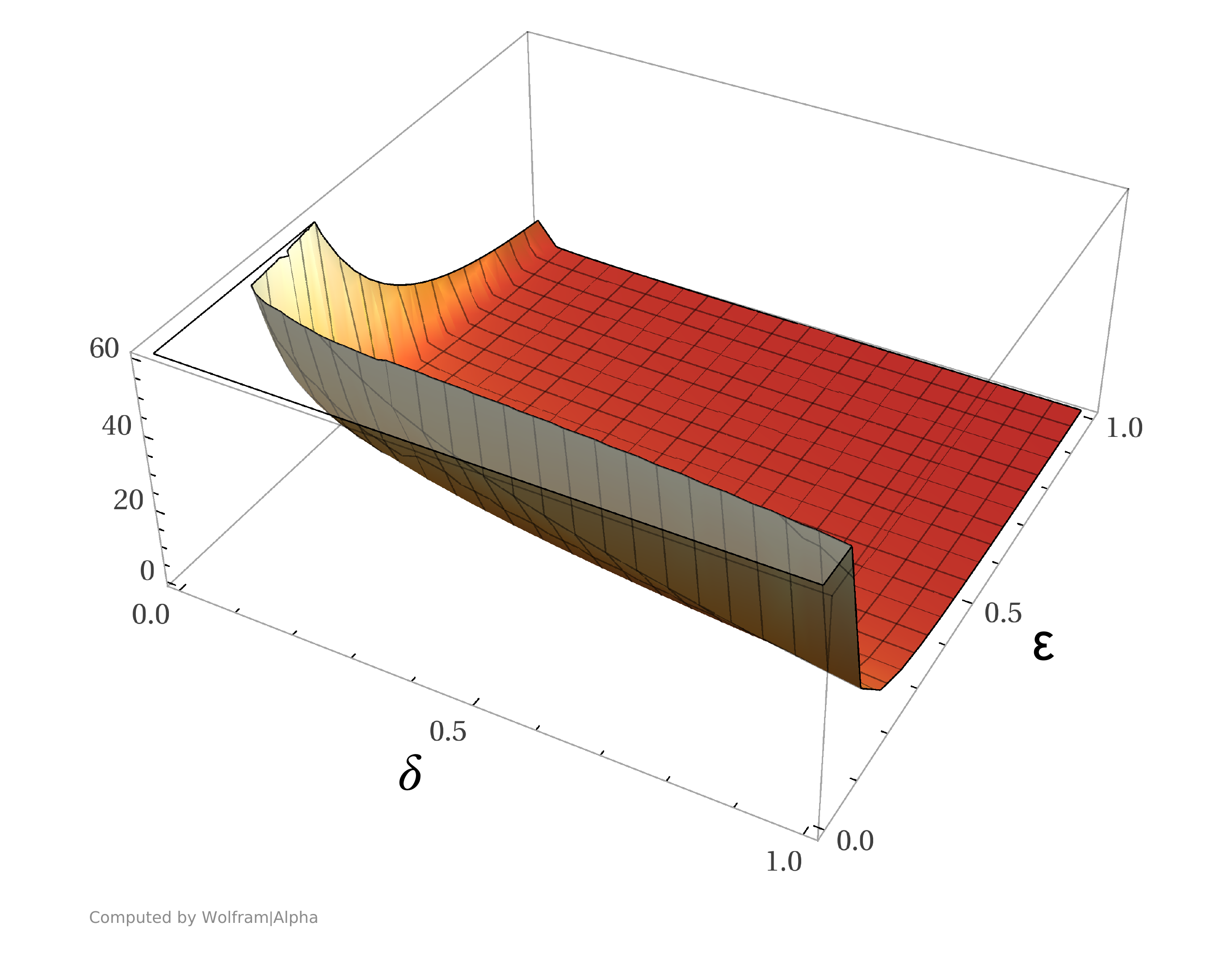}
\includegraphics[width=0.42\columnwidth]{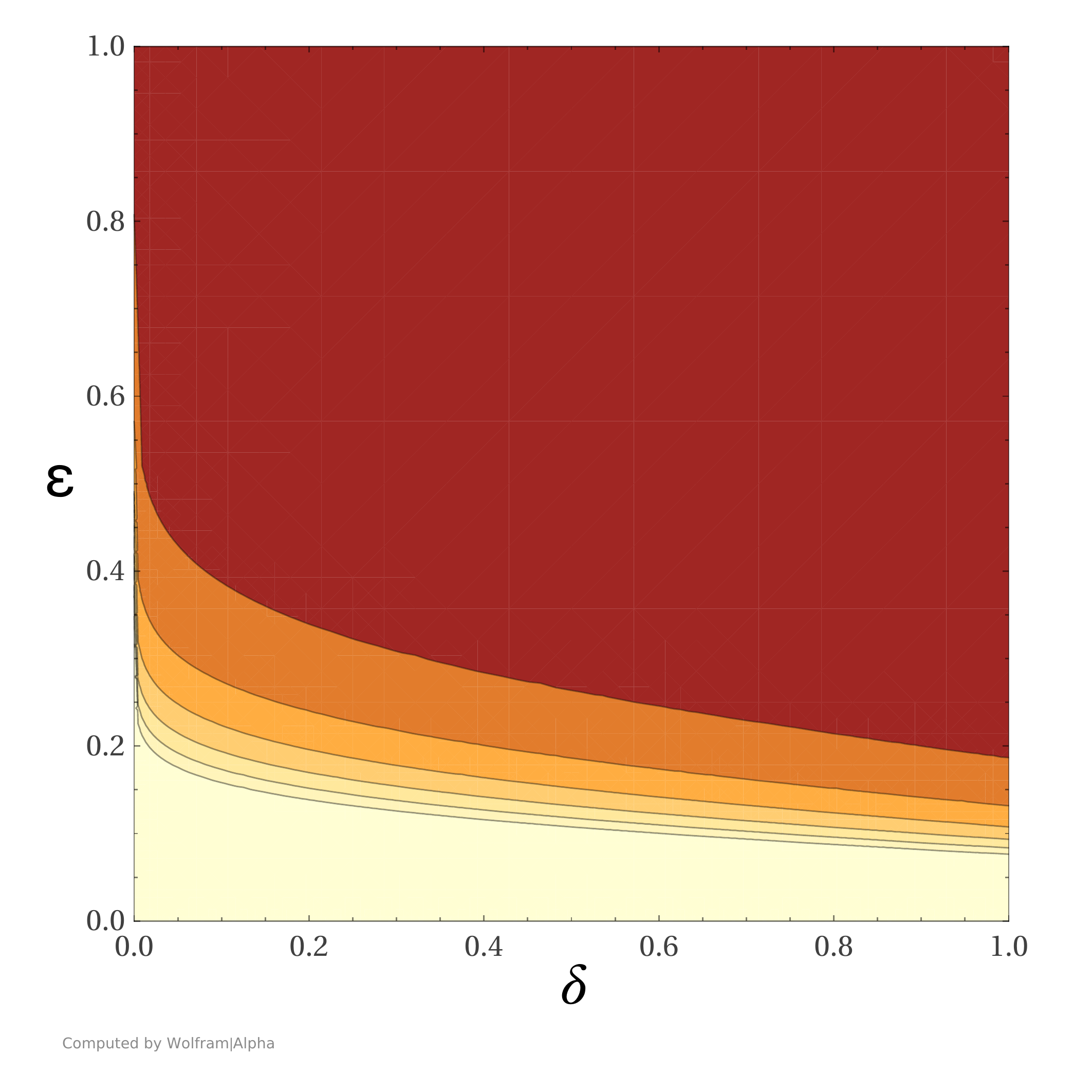}\\[-0.1cm]
\tiny\color{gray} 
Wolfram Alpha LLC. 2019. \scalebox{.5}{\url{https://www.wolframalpha.com/input/?i=plot+log(2/delta)/(2epsilon^2)+for+delta=0..1+epsilon=0..1} (access June 23, 2019).}
\caption{The number $n$ of Monte Carlo trials required, as confidence $\delta$ and accuracy $\epsilon$ vary. The surface plot (left) is a 3D plot of the function. The contour plot (right) visualizes how $n$ changes as $\delta$ and $\epsilon$ vary. The bright area at the bottom demonstrates substantial growth in $n$ as $\epsilon$ approaches 0.\vspace{-0.3cm}}
\label{fig:hfdplots}
\end{figure} 
\subsection{H\"{o}ffding's Inequality}
\hfd provides probabilistic bounds on the accuracy of an estimate $\hat \mu$ of the probability $\mu=P(\varphi)$ that a property $\varphi$ holds for an execution of a program.
The number of times $\varphi$ holds is a \emph{binomial random variable} $X=Bin(\mu,n)$. The probability $\mu$ that $\varphi$ holds is a \emph{binomial proportion}. An unbiased estimator of the binomial proportion is $\hat \mu=X/n$, i.e., the proportion of trials in which $\varphi$ was observed to hold. Given an \emph{accuracy parameter} $\epsilon>0$, H\"offding \cite{hoeff} provides an upper bound on the probability that the estimate $\hat \mu$ deviates more than $\epsilon$ from the true value $\mu$.\vspace{-0.05cm}
%{%\small
\begin{align}
P(|\hat \mu - \mu|\ge \epsilon) \le 2\exp(-2n\epsilon^2)\\[-0.55cm]\nonumber
\end{align}%}%
Given a \emph{confidence parameter} $\delta:0<\delta\le 1$, we can compute the number $n$ of sample executions required to guarantee the $(1-\delta)$-confidence interval $[\hat \mu-\epsilon,\hat \mu+\epsilon]$ as\vspace{-0.05cm}
%{%\small
\begin{align}
n>=\frac{\log(2/\delta)}{2\epsilon^2}\label{eq:hfdn}\\[-0.55cm]\nonumber
\end{align}%}%
Analogously, given $n$ Monte Carlo trials, with probability at least $1-\delta$, the absolute
difference between the estimate $\hat\mu$ and the true value $\mu$ is at most $\epsilon$ where\vspace{-0.2cm}
%{\small
\begin{align}
\epsilon \le \sqrt{\frac{\log(2/\delta)}{2n}}\label{eq:three}\\[-0.6cm]\nonumber
\end{align}%}%
Note that the probability that the absolute difference between estimate and true value exceeds our accuracy threshold $\epsilon$ \emph{decays exponentially} in the number of sample executions.

\subsection{Binomial Proportion Confidence Intervals}\label{sec:binomCIs}
\textbf{CP interval}. Clopper and Pearson \cite{clopper} discuss a \emph{statistical method}\footnote{A \emph{probabilistic method} starts with the underlying process and makes predictions about the observations the process generates. In contrast, a \emph{statistical analysis} starts with the observations and attempts to derive the parameters of the underlying random process that explain the observations. In other words, a statistical method requires observations to compute the estimates while a probabilistic method provides a-priori bounds on such estimates (e.g., based on Ro3 or HfD).} to compute a $(1-\delta)$-confidence interval $CI=[p_L,p_U]$ for a binomial proportion $\mu$, such that $\mu\in CI$ with probability \emph{at least} $(1-\delta)$ for all $\mu:0<\mu<1$.
The Clopper-Pearson (CP) interval is constructed by inverting the equal-tailed test based on the binomial distribution. Given the number of times $X$ that property $\varphi$ has been observed to hold in $n$ MC trials (i.e., $X: 0\le X\le n$), the confidence interval 
\begin{align}
CI_\text{CP} = [p_L,p_U]\label{eq:cpinterval}
\end{align}
is found as solution to the constraint
{\small%
\begin{align}
\sum_{k=X}^n \binom{n}{k}p_L^k(1-p_L)^{n-k} = \frac{\delta}{2} \text{\quad\quad and \quad\quad}
\sum_{k=0}^X \binom{n}{k}p_U^k(1-p_U)^{n-k} = \frac{\delta}{2}\nonumber
\end{align}}%
where $\binom{n}{k}=\frac{n!}{k!(n-k)!}$ is the binomial coefficient; $p_L=0$ when $X=0$ while $p_U=1$ when $X=n$.
The \emph{radius} $\epsilon$ of the CP interval is
\begin{align}
\epsilon = \frac{p_U - p_L}{2}\label{eq:cpradius}
\end{align}
Given the same number of Monte Carlo trials, the radius $\epsilon$ of the CP interval is lower-bounded by the generalized rule-of-three and upper-bounded by \hfd. 
The CP interval is \emph{conservative}, i.e., the probability that $\mu\in CI$ may actually be higher than $(1-\delta)$, particularly for probabilities $\mu$ that are close to zero or one. 

\textbf{Wald interval}. The most widely-used confidence intervals for binomial proportions are approximate and \emph{cannot} be used for $(\epsilon,\delta)$-approximate quantiative analysis.
The CP interval is computationally expensive and conservative. Hence, the most widely-used CIs are based on the approximation of the binomial with the normal distribution. For instance, the well-known \emph{Wald interval} \cite{onlyapprox} is
\begin{align} 
CI_\text{Wald}=\left[\hat \mu - z_{\delta/2}\sqrt{\frac{\mu(1-\mu)}{n}},\hat \mu + z_{\delta/2}\sqrt{\frac{\mu(1-\mu)}{n}}\right] 
\end{align}
where $z_{\delta/2}$ is the $(1-\frac{\delta}{2})$ quantile of a standard normal distribution. However, as we observe in our experiments, the Wald interval has poor coverage propertoes, i.e., for small $\mu$, the nominal 95\%-confidence interval actually contains $\mu$ only with a 10\% probability.

\begin{algorithm}[t]
\caption{$(\epsilon,\delta)$-Approximate Quantitative Program Analysis (Fully-Polynomial Randomized Approximation Scheme)}\label{alg:mcpa_theory}
\renewcommand{\algorithmicrequire}{\textbf{Input:}}
\renewcommand{\algorithmicensure}{\textbf{return}}
\begin{algorithmic}[1]
%\REQUIRE Program $\mathcal{P}$, Inputs $T$ \textcolor{darkgray}{(where $|T|=n$)}
%\REQUIRE Properties $\Phi$, Confidence $\delta$
\REQUIRE Program $\mathcal{P}$, Binary property $\varphi$
\REQUIRE Accuracy $\epsilon$, Confidence $\delta$
\STATE Let $n=\log(2/\delta)/(2\epsilon^2)$
\STATE Let $m=1$
\REPEAT
  \STATE Execute $\mathcal{P}$, observe outcome of $\varphi$, and increment $m$
  \STATE Let $\epsilon'$ be the radius of the current CP interval (cf. Eq.~(\ref{eq:cpradius}))
\UNTIL{$(m==n)$ \textbf{or} $(\epsilon'\le \epsilon)$}
\STATE Let $\hat\mu = X/m$ where $X$ is the total frequency $\varphi$ has held
\ENSURE \emph{``We \textbf{estimate} the prob. that $\varphi$ holds in $\mathcal{P}$ is $\hat\mu$ and \textbf{guarantee} that the \emph{true} prob. $\mu\in\hat\mu\pm\epsilon$ with prob. at least $(1-\delta)$.''}
%\ENSURE Estimates $\hat \mu_i$; Accuracy $\epsilon$
\end{algorithmic}
\end{algorithm}

\textbf{Wilson score interval}. The radius of the Wald interval is unreasonably small when the estimand $\mu$ gets closer to zero or one. The Wilson score CI mitigates this issue and has become the recommended interval for binomial proportions \cite{onlyapprox}. The \emph{Wilson score} interval \cite{wilson} is
\begin{align}
CI_\text{Wilson} = [\hat\mu' - \epsilon', \hat\mu' + \epsilon']
\end{align}
where $\hat\mu'$ is the relocated center estimate
{\small%
\begin{align}
\hat\mu' = \left.\left(\hat\mu + \frac{z^2_{\delta/2}}{2n}\right)\right/\left(1+\frac{z^2_{\delta/2}}{n}\right)
\end{align}}%
and $\frac{\epsilon'}{z_{\delta/2}}$ is the corrected standard deviation
{\small%
\begin{align}
\epsilon' = z_{\delta/2}\left.\sqrt{\frac{\hat\mu(1-\hat\mu)}{n}+\frac{z^2_{\delta/2}}{4n^2}}\right/\left(1+\frac{z^2_{\delta/2}}{n}\right)
\end{align}}%

\vspace*{-0.5cm}
\textbf{Evaluation}. We experimentally investigate properties of the three confidence intervals for binomial proportions in \autoref{sec:experiments}. We find that \emph{only} the Clopper-Pearson confidence interval guarantees that $\mu\in CI$ with probability \emph{at least} $(1-\delta)$ for all \mbox{$\mu:0\le\mu\le 1$}. The other two intervals \emph{cannot} be used for $(\epsilon,\delta)$-approximate quantiative analysis.
\vspace{-0.1cm}

\begin{algorithm}[t]
\caption{$(\epsilon,\delta)$-Approximate Quantitative Program Analysis (Reduced number of Clopper-Pearson CI calculations)}\label{alg:mcpa_practice}
\renewcommand{\algorithmicrequire}{\textbf{Input:}}
\renewcommand{\algorithmicensure}{\textbf{return}}
\begin{spacing}{0.88}
\begin{algorithmic}[1]
%\REQUIRE Program $\mathcal{P}$, Inputs $T$ \textcolor{darkgray}{(where $|T|=n$)}
%\REQUIRE Properties $\Phi$, Confidence $\delta$
\REQUIRE Program $\mathcal{P}$, Binary property $\varphi$
\REQUIRE Accuracy $\epsilon$, Confidence $\delta$
\STATE Let $n = n_\text{next} = \log(\delta) / \log(1-\epsilon)$
\FOR{each of $n_\text{next}$ Monte Carlo trials}
  \STATE Execute $\mathcal{P}$ and observe the outcome of $\varphi$
\ENDFOR
\STATE Let $\hat\mu = X/n$ where $X$ is the total frequency $\varphi$ has held
\IF{$0<\hat\mu<1$}
  \REPEAT
    \STATE Let $n_\text{next}=\left\lceil (z^2p_0 q_0 + z\sqrt{z^2p^2_0 q^2_0 + 2\epsilon p_0 q_0} + \epsilon)/\left(2\epsilon^2\right)\right\rceil$\\
    \quad\quad\quad\quad \ \ \ where $p_0=\hat\mu$ and $q_0=1-p_0$ and $z=z_{\delta/2}$
    \FOR{each of $n_\text{next}$ Monte Carlo trials}
      \STATE Execute $\mathcal{P}$ and observe the outcome of $\varphi$
    \ENDFOR
    \STATE Let $n = n + n_\text{next}$
    \STATE Let $\epsilon'$ be the radius of the current CP interval (cf. Eq.~(\ref{eq:cpradius}))
    \STATE Let $\hat\mu = X/n$ where $X$ is the total frequency $\varphi$ has held 
  \UNTIL{$\epsilon'\le \epsilon$}
\ENDIF

\ENSURE \emph{``We \textbf{estimate} the prob. that $\varphi$ holds in $\mathcal{P}$ is $\hat\mu$ and \textbf{guarantee} that the \emph{true} prob. $\mu\in\hat\mu\pm\epsilon$ with prob. at least $(1-\delta)$.''}
%\ENSURE Estimates $\hat \mu_i$; Accuracy $\epsilon$
\end{algorithmic}
\end{spacing}
\end{algorithm}

\subsection{\hspace{-0.1cm}Quantitative Monte Carlo Program Analysis}
Algorithm~\ref{alg:mcpa_theory} shows the procedure of the MCPA. \hfd provides the upper bound on the number of Monte Carlo trials required for an $(\epsilon, \delta)$-approximation of the probability $\mu$ that property $\varphi$ holds in program $\mathcal{P}$ (Eq.~(\ref{eq:hfdn})). The conservative Clopper-Pearson confidence interval provides an early exit condition (Line~6).

\textbf{Efficiency}. Algorithm~\ref{alg:mcpa_theory} runs in time that is polynomial in $1/\epsilon$ and in $\log(1/\delta)$ and is thus a \emph{fully-polynomial randomized approximation scheme} (FPRAS). The worst-case running time is visualized in \autoref{fig:hfdplots}. Reducing $\delta$ (i.e., increasing confidence) by an order of magnitude less than doubles execution time while reducing $\epsilon$ (i.e., increasing accuracy) by an order of magnitude increases execution time by two orders of magnitude.
However, computing the Clopper-Pearson (CP) CI after each Monte Carlo trial is expensive. In our experiments, computing the CP interval $10^5$ times takes 58 seconds, on average. Computing the CP interval for each of $n=2\cdot 10^{11}$ test inputs that OSS-Fuzz generates on a good day (cf. \autoref{fig:realworld}) would take about 3.8 years.

\textbf{Optimization}.
Hence, Algorithm~\ref{alg:mcpa_practice} predicts the number of Monte Carlo trials needed for a Clopper-Pearson interval with radius $\epsilon$. The estimator in Line~9 was developed by Krishnamoorthy and Peng \cite{predictclopper} and requires an initial guess $p_0$ of $\mu$. The algorithm computes the first guess $p_0=\hat\mu_L$ after running the minimal number of trials required for the given confidence and accuracy parameters (\mbox{Lines 1--6})---as provided by the generalized rule-of-three (Line~1). Subsequent guesses are computed from the improved maximum likelihood estimate (Line~14).

\section{Approximate Patch Verification}\label{sec:patchverif}

%https://research.fb.com/wp-content/uploads/2018/05/from-start-ups-to-scale-ups-opportunities-and-open-problems-for-static-and-dynamic-program-analysis.pdf?
\result{\emph{``It turns out that detecting whether a crash is fixed or not is an interesting challenge, and one that would benefit from further scientific investigation by the research community. [..] How long should we wait, while continually observing no re-occurrence of a failure (in testing or production) before we claim that the root cause(s) have been fixed?'' \cite{open2}\vspace{-0.1cm}}}\vspace{-0.1cm}

\begin{algorithm}[t]
\caption{Approximate Patch Verification}\label{alg:mcpa3} 
\renewcommand{\algorithmicrequire}{\textbf{Input:}}
\renewcommand{\algorithmicensure}{\textbf{return}}
\begin{algorithmic}[1]
%\State \textbf{Input}
%\State
\REQUIRE Program $\mathcal{P}_\text{fix}$, Binary property $\varphi$, Confidence $\delta$
\REQUIRE Total $n_\text{bug}$ and unsuccessful trials $X_\text{bug}$ in $\mathcal{P}_\text{bug}$
\STATE Let $n_\text{fix}=\log(\delta)/\log\left(1-p_L\right)$ where $p_L$ is the lower limit\\\quad\quad\quad\quad \ of the Clopper-Pearson interval (cf. Eq.~(\ref{eq:cpinterval})).
\FOR{Each of $n_\text{fix}$ Monte Carlo trials}
  \STATE Execute program $\mathcal{P}_\text{fix}$ and observe outcome of $\varphi$
  \STATE \textbf{if} $\varphi$ does \emph{not} hold in the current execution
  \STATE \textbf{then return} \emph{``Property $\varphi$ violated for current execution $\mathcal{P}_\text{fix}$''} 
\ENDFOR
\ENSURE \emph{``The null hypothesis can be rejected at significance-level at least $\delta$, to accept the alternative that failure rate has decreased.''}
%\ENSURE Estimates $\hat \mu_i$; Accuracy $\epsilon$
\end{algorithmic}
\end{algorithm}

Let $\mu_\text{bug}$ and $\mu_\text{fix}$ be the probability to expose an error before and after the bug was fixed, respectively. Suppose, we have $n_\text{bug}$ executions of the buggy program out of which $X_\text{bug}=Bin(\mu_\text{bug},n_\text{bug})$ exposed the bug. Hence, an unbiased estimator of $\mu_\text{bug}$ is $\hat\mu_\text{bug}=X_\text{bug}/n_\text{bug}$. We call an execution as \emph{successful} if no bug (of interest) was observed.
Given a confidence parameter $\delta:0<\delta< 1$, we ask how many successful executions $n_\text{fix}$ of the fixed program (i.e., $0=Bin(\mu_\text{fix},n_\text{fix})=X_\text{fix}$) are needed to reject the null hypothesis with statistical significance $\alpha=\delta$?

To \emph{reject the null hypothesis} at significance-level $\delta$, we require that there is \emph{no} overlap between both $(1-\delta)$-confidence intervals. For the buggy program version, we can compute the Clopper-Pearson interval $CI_\text{CP}=[p_L,p_U]$ (cf. Sec.~\ref{sec:binomCIs}). Recall that the probability that a property $\varphi$ holds is simply the complement of probability that $\not\varphi$ holds. For the fixed program version, we leverage the generalized rule-of-three $CI_\text{Ro3}=[0,1-\delta^{1/n_\text{fix}}]$ (cf. Eq.~(\ref{eq:nine})). In order to reject the null, we seek $n_\text{fix}$ such that\vspace{-0.1cm}
\begin{align}
1 - \delta^{\frac{1}{n_\text{fix}}} &< p_L\\[-0.3cm]\nonumber
\end{align}\vspace{-0.4cm}%
which is true for\vspace{-0.2cm}
\begin{align}
n_\text{fix} &> \frac{\log(\delta)}{\log\left(1-p_L\right)}
\end{align}
where $p_L$ is the lower limit of the Clopper-Pearson CI (cf. Eq.~(\ref{eq:cpinterval})).
 
%%%%
%If you compute hypothesis test from std error: 
% http://www.cscu.cornell.edu/news/statnews/Stnews73insert.pdf
% https://www.wolframalpha.com/input/?i=loglogplot+log(0.05)%2Flog(1%2F2+(2+-+(sqrt(2)+sqrt(2+n+%CE%BC%5E2+-+log(2%2F0.05)))%2Fsqrt(n)))+and+log(0.05)%2Flog(1-mu%2Bsqrt(log(2%2F0.05)%2F(2*n)))+where+n%3D10%5E9+from+mu%3D0.00001..1
%%%

\begin{figure}\centering
\includegraphics[width=0.52\columnwidth, clip=true,trim= 0 0 0 0.8cm]{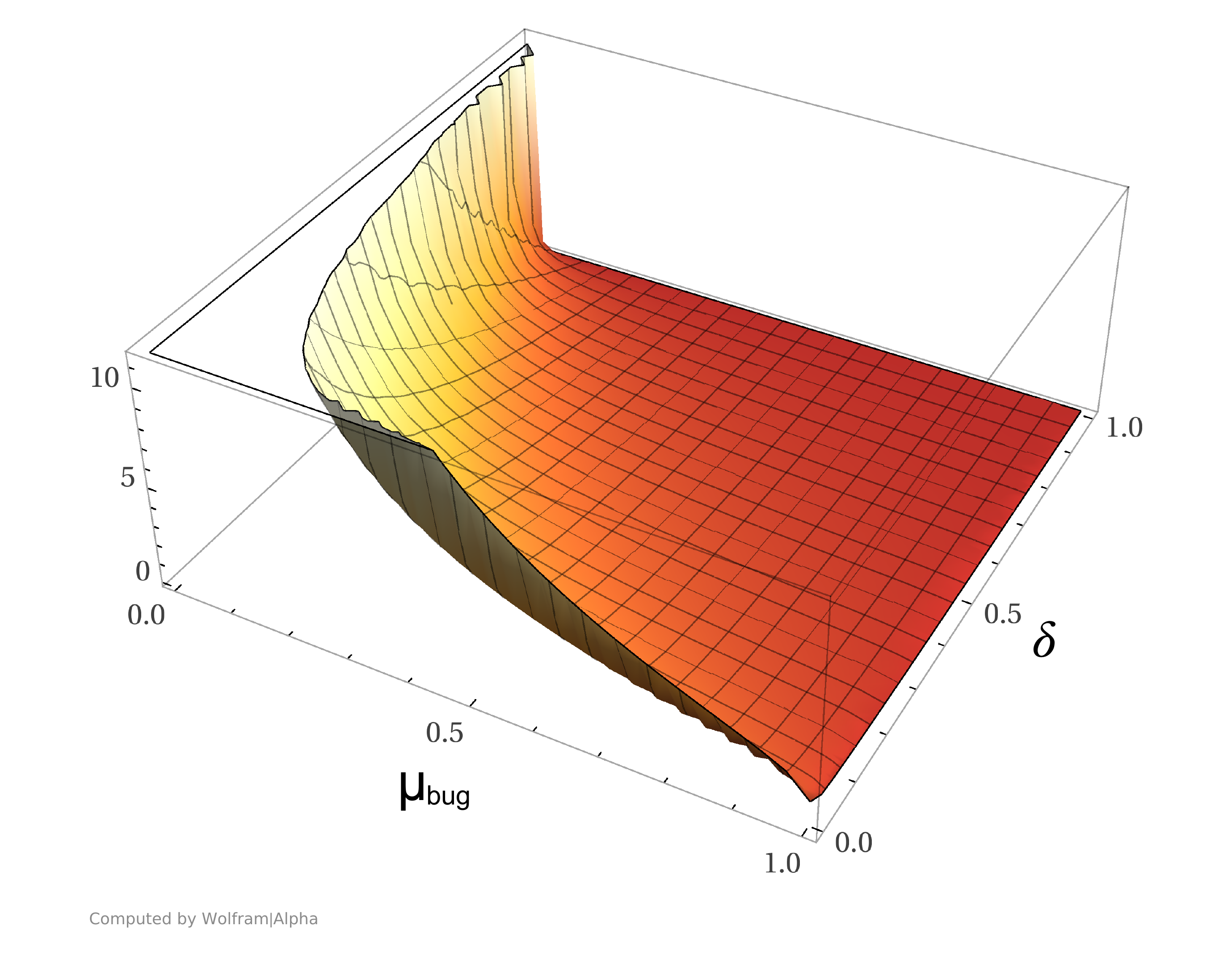}
\includegraphics[width=0.42\columnwidth]{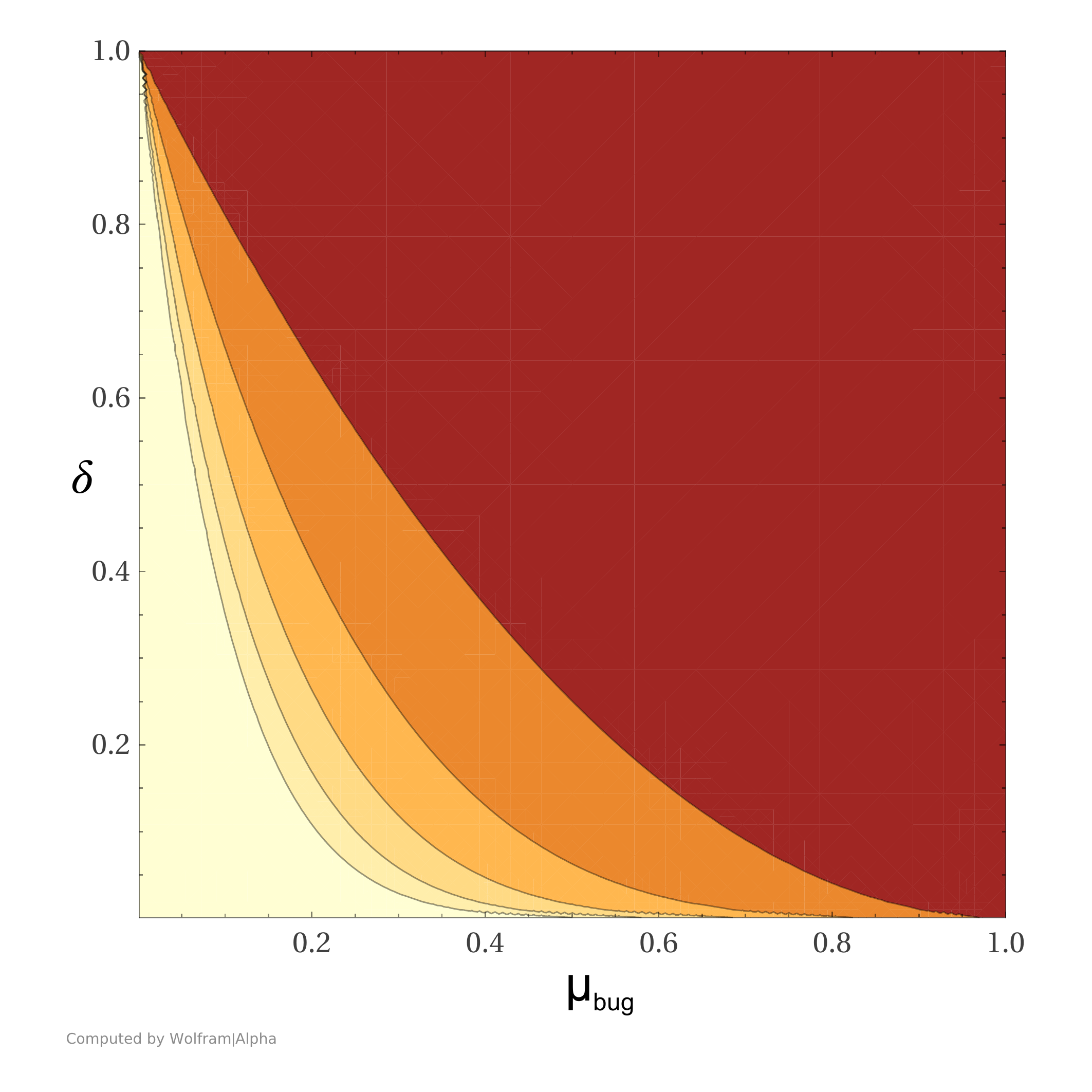}\\[-0.1cm]
\tiny\color{gray} 
Wolfram Alpha LLC. 2019. \scalebox{.5}{\url{https://www.wolframalpha.com/input/?i=plot+log(delta)\%2Flog(1-mu\%2Bsqrt(log(2\%2Fdelta)\%2F(2*4e6)))+where+mu=0.001..1,delta=0.001..1} (access June 23, 2019)}.
\caption{The number $n_\text{fix}$ of Monte Carlo trials required for the patched program to reject the Null and conclude that the failure rate has indeed decreased as significance-level $\delta$ and failure rate $\hat\mu_\text{bug}$ for the buggy program vary ($n_\text{bug}=4\text{e}6$). The contour plot shows how $n_\text{fix}$ changes as $\delta$ and $\mu_\text{bug}$ vary.\vspace{-0.1cm}}  
\label{fig:apvplots}
\end{figure}

\textbf{Efficiency}. The CP limit $p_L$ is upper- and lower-bounded by the generalized rule-of-three and \hfd, respectively, 
\begin{align}
\hat\mu_\text{bug}-\sqrt{\frac{\log(2/\delta)}{2n_\text{bug}}} \le  p_L \le \hat\mu-\left(1-\delta^{\frac{1}{n_\text{bug}}}\right) 
\end{align}
for all $0\le \mu_\text{bug} \le 1$. Hence, Algorithm \ref{alg:mcpa3} is an FPRAS that runs in time that is polynomial in $\log(1/(1-\hat\mu_\text{bug}))^{-1}$ and in $\log(1/\delta)$. The \emph{worst-case efficiency} of Algorithm~\ref{alg:mcpa3} is visualized in \autoref{fig:apvplots}.
For instance, if 800k of 200 \emph{billion} executions expose a bug in $\mathcal{P}_\text{bug}$, then it requires less than 1.8k executions to reject the Null at significance level $p<0.001$ in favor of the alternative that the probability of exposing a bug has indeed reduced for $\mathcal{P}_\text{fix}$.

%\newpage
\section{Experiments}\label{sec:experiments}
We implemented Algorithms~\ref{alg:mcpa2}--\ref{alg:mcpa3} into 300 lines of \texttt{R} code. The \texttt{binom} package \cite{binom} was used to compute various kinds of confidence intervals. \texttt{R} is a statistical programming language.\vspace{-0.1cm}

\subsection{Approximate Quantitative Analysis}
\begin{figure}[h]
\vspace{-0.2cm}
\includegraphics[width=\columnwidth]{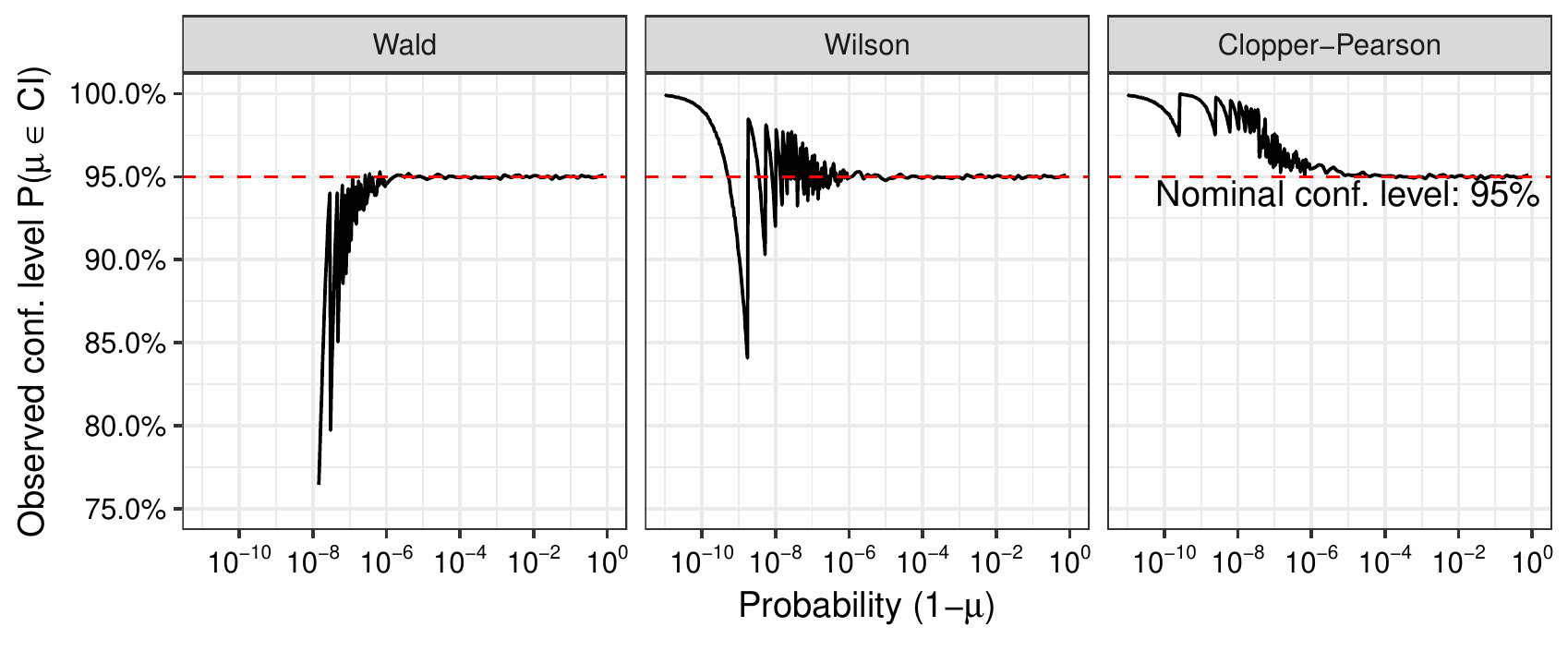}%
\vspace{-0.3cm}
\caption{The observed probability that $\mu\in CI$ given the nominal probability $(1-\delta)=95\%$ as $\mu$ varies. We generated 100 \emph{thousand} repetitions of 100 \emph{million} trials for one \emph{thousand} values of $\mu$, i.e., $\approx 10^{18}$ Monte Carlo trials in total.}
\label{fig:rq1} 
\end{figure}
\textbf{RQ1. Can the Clopper-Pearson interval $CI_\text{CP}$ be used for $(\epsilon,\delta)$-approximate quantitative analysis?} An $(\epsilon,\delta)$-approx\-i\-ma\-tion $\hat\mu$ of a binomial proportion $\mu$ guarantees that $\mu\in[\hat\mu-\epsilon,\hat\mu+\epsilon]$ with probability \emph{at least} $(1-\delta)$. \autoref{fig:rq1} shows the results for $10^{18}$ simulation experiments. For various values of $\mu$, we generated $n=10^8$ trials by sampling from $\mu$ and computed the 95\% confidence interval according to the methods of Wald, Wilson, and Clopper-Peason. We repeated each experiment $10^5$ times and measured the proportion of intervals that contain $\mu$. This proportion gives the observed while 95\% is the nominal confidence-level.\vspace{0.1cm}

\result{\emph{Yes. The Clopper-Pearson 95\%-confidence interval $CI_\text{CP}$ contains $\mu$ with probability \emph{at least} 95\% for all values of $\mu$.}
Algorithms~\ref{alg:mcpa_theory} and \ref{alg:mcpa_practice} provide valid $(\epsilon,\delta)$-approx\-i\-ma\-tions. However, both the Wald and the Wilson procedures provide 95\%-CIs that contain~$\mu$ with probability \emph{less than} 95\%, especially for $\mu\to 1$ (or $\mu\to 0$, resp.). For instance, if $\mu=1-10^{-8}$, the Wald 95\%-confidence interval $CI_\text{Wald}$, which is the most widely-used interval for binomial proportions, contains $\mu$ with probability $P(\mu\in CI_\text{Wald})<80\%$.}

\begin{figure}[h]
\includegraphics[width=\columnwidth]{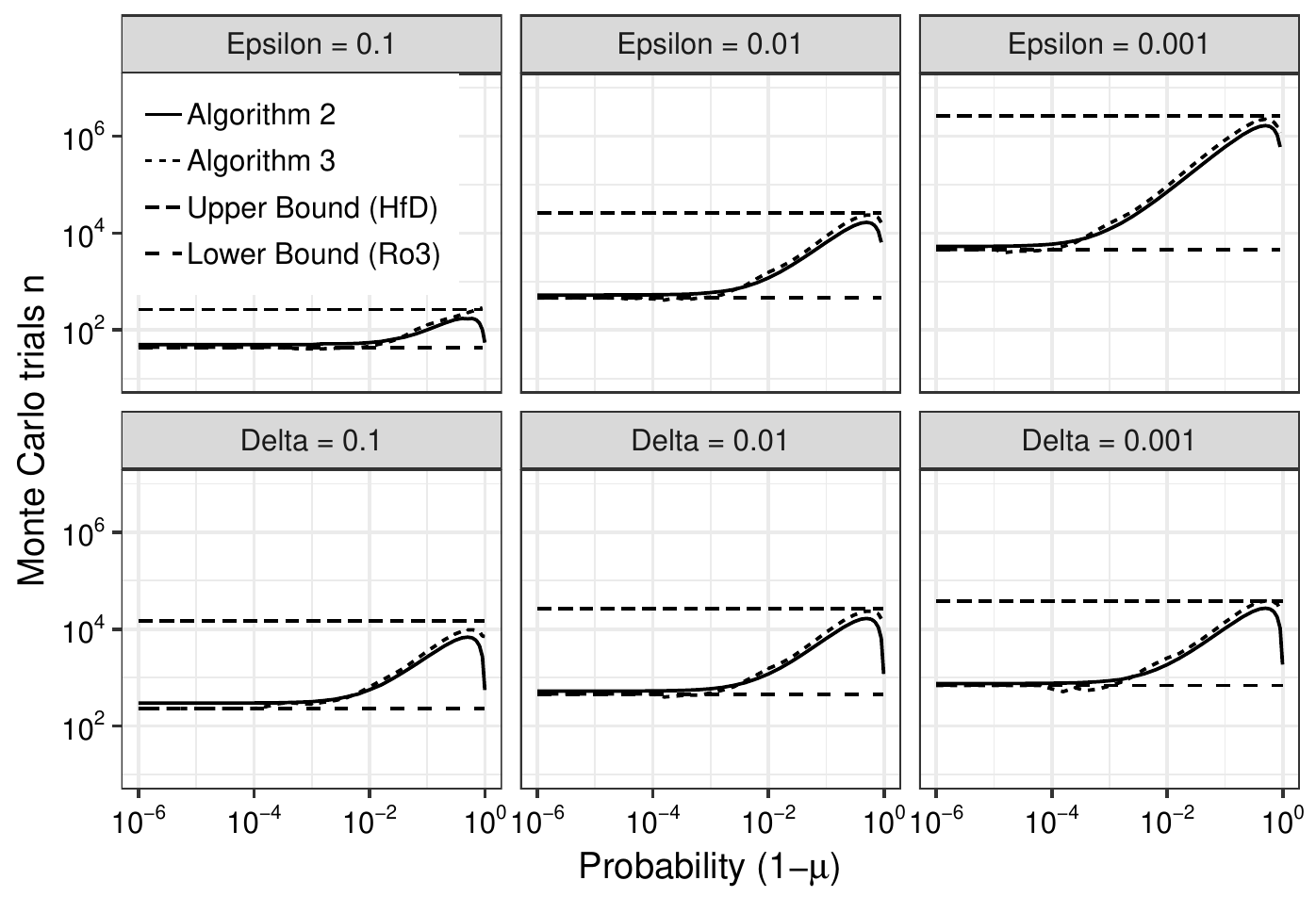}%
\vspace{-0.3cm}
\caption{Required MC trials $n$ for an $(\epsilon,\delta)$-approximation of $\mu$. The  rule-of-three (Ro3) provides the lower while \hfd (HfD) provides the upper bound for all $0\le\mu\le 1$. In the first row, $\delta=0.01$. In the second row, $\epsilon=0.01$.\vspace{-0.3cm}}  
\label{fig:rq2} 
\end{figure}
\textbf{RQ2 (Efficiency). How efficient are Algorithms~\ref{alg:mcpa_theory} and \ref{alg:mcpa_practice} w.r.t. the probabilistic upper and lower bounds?} Here, \emph{efficiency} is measured as the number of MC trials required to compute an $(\epsilon,\delta)$-approximation of the probability $\mu$ that $\varphi$ holds. The upper and lower bounds for all $\mu:0\le\mu\le 1$ are given by \hfd (HfD) and the generalized rule-of-three (Ro3), respectively. \autoref{fig:rq2} shows the efficiency of both algorithms as $\mu$ varies, as well as the probabilistic bounds, for various values of $\epsilon$ and $\delta$. Each experiment was repeated 1000 times and average values are reported.

\result{\emph{Both algorithms approach the probabilistic lower bound (Ro3) as $\mu\to 1$ (or $\mu\to 0$, resp.).} For instance, for $\delta=0.01$, $\epsilon=0.001$, at least 4603 and most 2.6 \emph{million} Monte Carlo trials are required; for $\mu=10^{-4}$, Algorithm~\ref{alg:mcpa_theory} requires 4809 trials, which is just 4\% above the lower bound. Moreover, \emph{Algorithm~\ref{alg:mcpa_practice} requires only slightly more Monte Carlo trials than Algorithm~\ref{alg:mcpa_theory} but reduces the number of computed intervals substantially from $n$ to at most 2.}}\vspace{-0.2cm}

\begin{figure}[h]\footnotesize
\begin{tabular}{rrl||rrr|rr}
\multicolumn{3}{c||}{\textbf{BinaryTree}} &\multicolumn{5}{c}{\textbf{BinomialHeap}}\\
& \multicolumn{2}{c||}{$v\in[0,9]$} & & \multicolumn{2}{c|}{$v\in[0,9]$} &\multicolumn{2}{c}{$v\in[0,499]$}\\%\cline{5-8}
\texttt{Loc} & \multicolumn{1}{c}{$\hat\mu$} & \multicolumn{1}{c||}{$\mu$} & \texttt{Loc} & \multicolumn{1}{c}{$\hat\mu$} & \multicolumn{1}{c|}{$\mu$} & \multicolumn{1}{c}{$\hat\mu$} & \multicolumn{1}{c}{$\mu$}\\\hline
0 & 0.9374 & 0.9375 & 1 & 0.64464 & 0.64451 & 0.74833 & 0.74799\\
1 & 0.4589 & 0.4592 & 4 & 0.22842 & 0.22831 & 0.31051 & 0.31062	\\
2 & 0.5740 & 0.5745 & 6 & 0.60359 & 0.60350 & 0.68618 & 0.68599\\
3 & 0.4591 & 0.4592 & 7 & 0.26947 & 0.26932 & 0.37265 & 0.37263\\
4 & 0.5742 & 0.5745 & 8 & 0.22842 & 0.22831 & 0.31051 & 0.31062\\
8 & 0.0202 & 0.0203 & 9 & 0.38148 & 0.38161 & 0.40648 & 0.40583\\
9 & 0.0189 & 0.0189 & 10 & 0.30421 & 0.30389 & 0.40398 & 0.40416\\
10 & 0.0347 & 0.0346 & 13 & 0.00767 & 0.00765 & 0.00025 & 0.00025\\
11 & 0.0361 & 0.0361 & 14 & 0.11879 & 0.11863 & 0.00275 & 0.00274\\
12 & 0.0361 & 0.0361 & 15 & 0.00767 & 0.00765 & 0.00025 & 0.00025\\
13 & 0.1077 & 0.1077 & 16 & 0.05551 & 0.05552 & 0.00150 & 0.00149\\
14 & 0.5742 & 0.5745 & 17 & 0.06640 & 0.06621 & 0.00138 & 0.00137\\
15 & 0.5739 & 0.5745 & 18 & 0.04784 & 0.04787 & 0.00125 & 0.00124\\
& &                  & 19 & 0.00454 & 0.00456 & 0.00012 & 0.00012\\
& &                  & 20 & 0.00767 & 0.00765 & 0.00025 & 0.00024\\
%& &                  & 21 & 0.02604 & 0.02610 & 0.00075 & 0.00075\\\hline
\end{tabular}
\vspace{-0.2cm}
\caption{Branch probabilities computed by Algorithm~\ref{alg:mcpa_practice} ($\hat\mu$) and JPF-ProbSym ($\mu$).  We generated sufficient MC trials to guarantee ($\epsilon,\delta)$-approximations for \emph{all} $0\le\mu\le 1$ s.t. $\delta=\epsilon=10^{-3}$. That is, $3.8\cdot 10^6$ executions of random values $v$ on BinaryTree (4.6 \emph{seconds}) and BinomialHeap (0.5 \emph{seconds}).}
\label{fig:rq3} 
\end{figure}

\textbf{RQ3 (Experience with PSE)}. Geldenhuys et al. \cite{pse} developed what is now the state-of-the-art of exact quantitative program analysis, \emph{probabilistic symbolic execution} (PSE). The authors implemented PSE into the JPF-ProbSymb tool and evaluated it using the two programs BinaryTree and BinomialHeap. For both programs, \autoref{fig:rq3} shows the exact branch probabilities $\mu$ as computed by PSE and the corresponding $(\epsilon,\delta)$-approximations as computed by Algorithm~\ref{alg:mcpa_practice}. To generate a Monte Carlo trial, we randomly sampled from the same domain that PSE is given.

As we can see, our Monte Carlo program analysis (MCPA) approximates the branch probabilities $\mu$ within the specified minimum accuracy. For $\delta=\epsilon=0.001$, our estimates $\hat\mu$ are usually within $\pm 0.0005$ of the true values $\mu$. We would like to highlight that \emph{approximate PSE} \cite{filieri2,filieri3,filieri4,filieri5} does not provide any such guarantees on the accuracy of the produced estimates. We also observed that PSE is tractable only for very small integer domains (e.g,. $v\in [0,9]$ \cite{pse} or $v\in [0,100]$ \cite{bihuan}) while our MCPA works with arbitrarily large domains just as efficiently. In fact, unlike PSE, MCPA works for executions generated \emph{while} the software is deployed and used.

\result{\emph{The results of Alg.~\ref{alg:mcpa_practice} are comparable to those produced by JPF-ProbSym with an error $\epsilon$ that can be set arbitrarily close to zero.}}

However, without the additional machinery of PSE, our MCPA algorithm produces the analysis result much faster. For BinomialHeap, our algorithm took half a second (0.5s) while JPF-ProbSymb took 57 seconds. For BinaryTree, our algorithm took about 5 seconds while JPF-ProbSymb took about seven \emph{minutes}. In general, we expect that a sufficiently large number of Monte Carlo trials can be generated quickly for arbitrarily large programs (cf.~\autoref{fig:realworld} on page~\pageref{fig:realworld}), which makes MCPA effectively scale-oblivious.

\result{\emph{Versus JPF-ProbSym, our MCPA is orders of magnitude faster}.}

\subsection{Approximate Software Verification}
\begin{figure}[h]
\vspace{-0.2cm}
\includegraphics[width=\columnwidth]{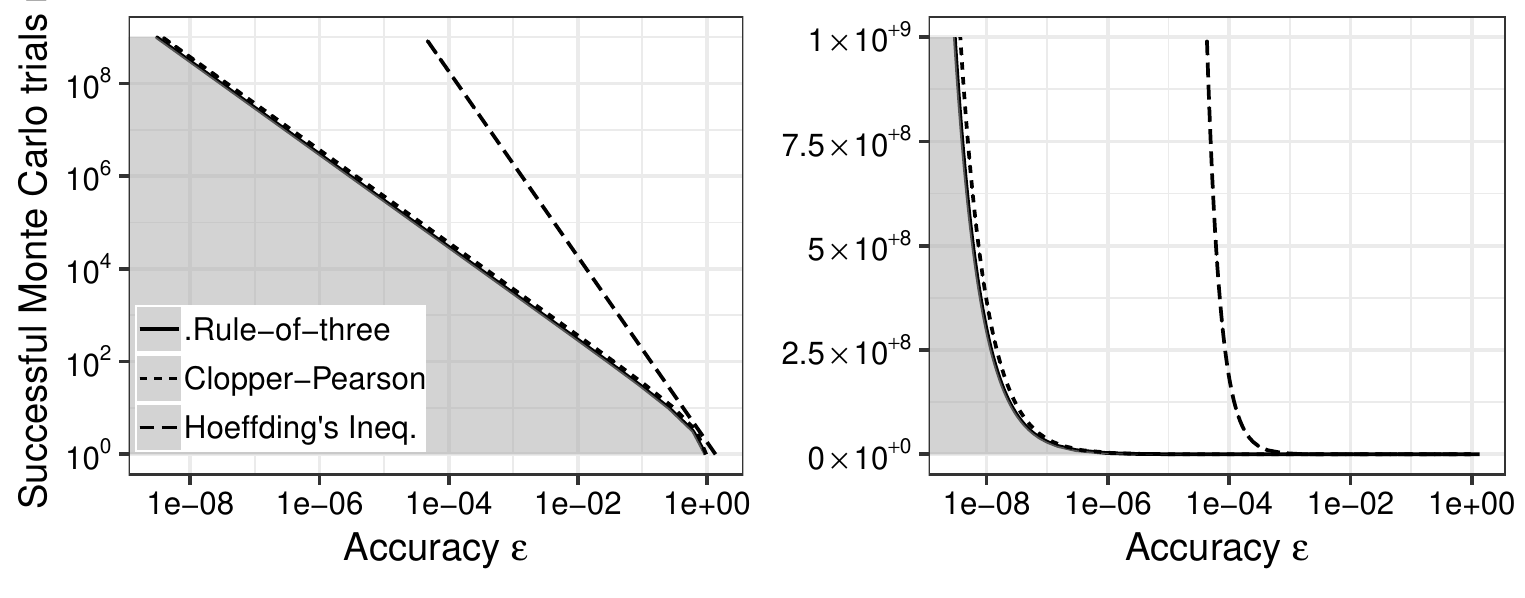}%
\vspace{-0.3cm}    
\caption{Efficiency versus residual risk. Both plots show the number of required Monte Carlo trials $n$, where $\varphi$ is observed to hold, to \emph{guarantee}---with an error that exceeds $\epsilon$ with probability at most $\delta=0.05$---\emph{that $\varphi$ always holds}.}
\label{fig:rq4}  
\end{figure}

\textbf{RQ4 (Residual Risk). For a given allowable residual risk $\epsilon$, how efficient is Alg.~\ref{alg:mcpa2} in providing the probabilistic guarantee that no bug exists when none has been observed?} \autoref{fig:rq4} demonstrates the efficiency of approximate software verification. Alg.~\ref{alg:mcpa2} exhibits a performance that is nearly inversely proportional to the given allowable residual risk. For instance, it requires about $3\cdot 10^5$ trials where no error is observed to guarantee that no error exists with an error that exceeds a residual risk of $\epsilon=10^{-5}$ with probability at most 5\%. Decreasing $\epsilon$ by an order of magnitude to $\epsilon=10^{-6}$ also increases the number of required trials only by an order of magnitude to $3\cdot 10^{-6}$.

\result{\emph{Algorithm~\ref{alg:mcpa2} is highly efficient. It exhibits a performance that is nearly inversely proportional to the given allowable residual risk.}}   

Moreover, the Clopper-Pearson interval $CI_\text{CP}$ provides an upper limit that is very close to the probabilistic lower bound as given by the rule-of-three---confirming our observation in RQ2.

\subsection{Approximate Patch Verification}
\begin{figure}[h]
\vspace{-0.2cm}
\includegraphics[width=\columnwidth,clip,trim=0 4cm 0 4cm]{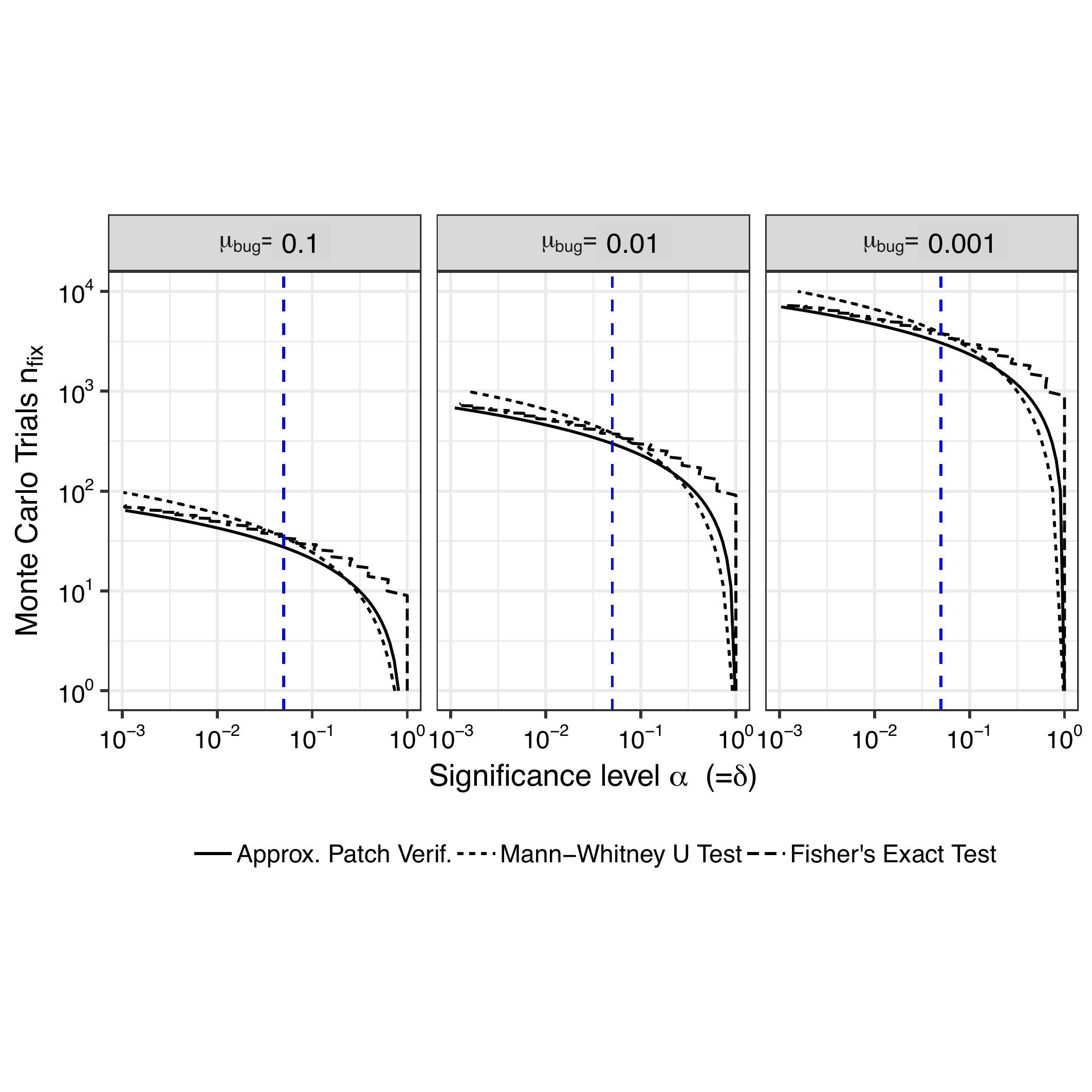}
\vspace{-0.6cm}  
\caption{Efficiency versus significance. The number of required successful executions $n_\text{fix}$ of $\mathcal{P}_\text{fix}$ to reject the Null at significance $\alpha=\delta$. The blue, dashed line shows $\alpha=0.05$. We conducted $10^{15}$ simulation experiments, in total; i.e., $10^3$ repetitions of $n_\text{bug}=10^6$ trials on $\mathcal{P}_\text{bug}$ for each of $10^2$ values of $n_\text{fix}\le 10^4$ on $\mathcal{P}_\text{fix}$, and $3$ values of $\mu_\text{bug}$.}
\label{fig:rq5}
\end{figure}

\textbf{RQ5 (Efficiency). How efficient is Algorithm~\ref{alg:mcpa3} in rejecting the null hypothesis at a desired p-value compared to existing hypothesis testing methodologies?} In \autoref{fig:rq5}, we compare our MCPA to the Fisher's exact test and the Mann-Whitney U test. The \emph{Fisher's exact test} is the standard two-sample hypothesis test for binomial proportions when the one of the estimates (here $\hat\mu_\text{fix}$) is zero or one.\footnote{We note that the popular Chi-squared test, while easier to compute, is an approximation method that \emph{cannot} be used when the one of the estimates $\hat\mu_\text{bug}$, $\hat\mu_\text{fix}$ is 0~or~1.}
The \emph{Mann-Whitney U test} is the standard test used in the AB testing platform at Netflix \cite{netflix,netflixStat} to assess, e.g., performance degradation between versions. The U test is a non-parametric test of the null hypothesis that it is equally likely that a randomly selected value from one sample will be less than or greater than a randomly selected value from a second sample. 

\result{\emph{Algorithm~\ref{alg:mcpa3} is more efficient than existing techniques for high levels of confidence (e.g, p-value $ =\alpha\le 0.05$) on this task of approximate patch verification. The Mann-Whitney test that Netflix uses for A/B testing performs worst on this task; the difference to our MCPA increases as $\alpha\to 0$. Fisher's exact test approaches the performance of our MCPA as $\alpha\to 0$.}}

\subsection{Threats to Validity}
As in every empirical study, there are threats to the validity of the results. 
In terms of \emph{external validity}, the assumptions of Monte Carlo program analysis (MCPA) may not be realistic. A particular advantage of simulation experiments is their generality: Our results apply to arbitrary software systems of arbitrary size, as long as the stochastic process of generating the Monte Carlo trials satisfies the assumptions of MCPA. Thus, whether or not the assumptions are realistic is the only threat to external validity. In \autoref{sec:mcpa}, we discuss these assumptions, how realistic they are, and mitigation strategies in case they do not hold. Particularly, most assumptions are shared with those of automatic software testing.

In terms of \emph{internal validity}, we cannot guarantee that our implementation is correct, but make our script available for the reader to scrutinize the implementation and reproduce our results.\footnote{\url{https://www.dropbox.com/sh/fxiz3ycvvbi51xg/AACk0UyO_nG0UlPMEH645jbXa?dl=0}} To minimize the probability of bugs, we make use of standard R packages, e.g., \texttt{binom} for the computation of the confidence intervals. 
\section{Related Work}\label{sec:related}
We are interested in the probability that a \emph{binary property}\footnote{Quantitative program properties, such as energy, are considered by checking whether they exceed a threshold (for an arbitrary number of thresholds).} $\varphi$ holds for an execution of a \emph{terminating} program. This problem is also tackled by probabilistic symbolic execution \cite{pse}. After a discussion of the relationship to probabilistic symbolic execution, we discuss differences to A/B testing and runtime verification techniques.

\textbf{Probabilistic Symbolic Execution} (PSE)
 computes for each path $i$ that satisfies the property $\varphi$ the probability $p_i$ using symbolic execution and exact \cite{pse,filieri1} or approximate model counting \cite{filieri2,filieri3,filieri4,filieri5}, and then computes $\mu=\sum_ip_i$ to derive the probability that $\varphi$ holds.
In contrast to MCPA, PSE allows to analyse non-terminating programs by fixing the maximum length of the analyzed path. However, in contrast to PSE, MCPA does not require a formal encoding of the operational distribution as usage profiles. Instead, the analysis results are directly derived from actual usage of the software system. 
While PSE is a static quantitative program analysis that never actually executes the program, MCPA is a dynamic quantitative analysis technique that requires no access to the source code or the heavy machinery of constraint encoding and solving. In contrast to approximate PSE \cite{filieri2,filieri3,filieri4,filieri5}, MCPA provides probabilistic guarantees on the accuracy of the produced point estimate $\hat\mu$. In contrast to statistical PSE \cite{filieri4}, MCPA does not require an exact model count for the branch probabilities of each \texttt{if}-conditional. Other differences to PSE are discussed in \autoref{sec:motivation}.%However, PSE is more useful than MCPA when analysing paths of finite length in non-terminating programs.

\textbf{A/B Testing} \cite{abtesting,netflix} provides a statistical framework to evaluate the performance of two software systems (often the new version versus the old version) within quantifiable error. In contrast to A/B testing, $(\epsilon,\delta)$-approximate quantitative analysis as well as $(\epsilon,\delta)$-approximate software verification pertains to a single program. Moreover, in simulation experiments, our $(\epsilon,\delta)$-approximate patch verification outperforms the Fisher's exact test and the Mann-Whitney U test, which is the two-sample hypothesis test of choice for A/B testing at Netflix \cite{netflixStat,netflix}. We also note that A/B testing is subject to the same assumptions.

\textbf{Statistical/Probabilistic Model Checking}.
Informally speaking, $(\delta,\epsilon)$-approximate software verification is to probabilistic and statistical model checking as software verification is to classical model checking. Our MCPA techniques are neither focused on temporal properties nor concerned with a program's state space. More specifically, instead of verifying the expanding prefix of an infinite path through the program's state space, MCPA requires several distinct, terminating executions (Sec.~\ref{sec:related}). In this setting, the work on probabilistic symbolic execution \cite{pse,filieri1,filieri2} is most related.

% https://www.lrde.epita.fr/dload/papers/darbon.06.iccp.pdf
In model checking, there is the problem of deciding whether a model $\mathcal{S}$ of a potentially \emph{non-terminating} stochastic system satisfies a temporal logic property $\phi$ with a probability greater than or equal to some threshold $\theta$: $\mathcal{S}\models P_{\ge \theta}(\phi)$. %The stochastic system $\mathcal{S}$ is modelled using, e.g., a discrete or continuous time markov chain. 
The temporal property $\phi$ is specified in a probabilistic variant of linear time logic (LTL) or computation tree logic (CTL).  
For instance, we could check: ``\emph{When a shutdown occurs, the probability of a system recovery being completed between 1 and 2 hours without further failure is greater than 0.75}'':\vspace{-0.1cm}
$$\mathcal{S}\models \text{down}\rightarrow P_{>0.75}[\neg\text{fail}\ U^{[1,2]}\ \text{up}]$$

\vspace*{-0.1cm}
Probabilistic model checking (PMC) \cite{pmc,probmodelcheck1,probmodelcheck2} checks such properties using a (harder-than-NP) analytical approach. Hence, PMC becomes quickly untractable as the number of states increases. PMC also requires a formal model of the stochastic system $\mathcal{S}$ as discrete- or continuous-time Markov chain. 
Statistical model checking (SMC) \cite{smc,sen} does not take an analytical but a statistical approach. SMC checks such properties by performing \emph{hypothesis testing} on a number of (fixed-length) simulations of the stochastic system $\mathcal{S}$. Similar to MCPA, Sen et al. \cite{sen} propose to \emph{execute} the stochastic system directly. However, Sen et al. assume that a ``trace'' is available, i.e., $\mathcal{S}$ can identify and report the sequence of states which $\mathcal{S}$ visits and how long each state transition takes. The properties that are checked concern particular state sequences and time intervals. In contrast, the focus of our current work is on binary (rather than temporal) properties for executions of terminating (rather than non-terminating) systems. In our setting, the concept of time is rather secondary. Moreover, $(\delta,\epsilon)$-approximate software and patch verification are just two particular instantiations of MCPA.

\section{Conclusion}\label{sec:conc}
We are excited about the tremendous advances that have been made in scalable program analysis, particularly in the area of separation logic. For instance, the ErrorProne static analysis tool is routinely used at the scale of Google's two-billion-line codebase \cite{errorprone}. The Infer tool has substantial success at finding bugs at Facebook scale \cite{infer}. Yet, there still remain several open challenges; e.g., the developers of Infer set the clear expectation that the tool may report many false alarms, does not handle certain language features, and can only report certain types of bugs.\footnote{\url{https://fbinfer.com/docs/limitations.html}}
We strongly believe that many of these challenges are going to be addressed in the future. However, we also feel that it is worthwhile to think about alternative approaches to scalable program analysis. Perhaps to think about approaches that do not attempt to maintain formal guarantees at an impractical cost. If we are ready to give up on soundness or completeness in favor of scalability, we should at least be able to quantify the trade-off.
The probabilistic and statistical methodologies that are presented in this paper represent a significant progress in this direction. 
%\input{sections/experiments.old.tex}

%% Acknowledgments
\begin{acks}                            %% acks environment is optional
                                        %% contents suppressed with 'anonymous'
  %% Commands \grantsponsor{<sponsorID>}{<name>}{<url>} and
  %% \grantnum[<url>]{<sponsorID>}{<number>} should be used to
  %% acknowledge financial support and will be used by metadata
  %% extraction tools.
%   This material is based upon work supported by the
%   \grantsponsor{GS100000001}{National Science
%     Foundation}{http://dx.doi.org/10.13039/100000001} under Grant
%   No.~\grantnum{GS100000001}{nnnnnnn} and Grant
%   No.~\grantnum{GS100000001}{mmmmmmm}.  Any opinions, findings, and
%   conclusions or recommendations expressed in this material are those
%   of the author and do not necessarily reflect the views of the
%   National Science Foundation.
We thank Prof. David Rosenblum for his inspiring ASE'16 keynote speech on probabilistic thinking \cite{probThinking}.
This research was partially funded by the Australian Government through an Australian Research Council Discovery Early Career Researcher Award (DE190100046).
\end{acks}


%% Bibliography
%\bibliography{references}

\begin{thebibliography}{52}

%%% the Web address.

\ifx \showCODEN    \undefined \def \showCODEN     #1{\unskip}     \fi
\ifx \showDOI      \undefined \def \showDOI       #1{#1}\fi
\ifx \showISBNx    \undefined \def \showISBNx     #1{\unskip}     \fi
\ifx \showISBNxiii \undefined \def \showISBNxiii  #1{\unskip}     \fi
\ifx \showISSN     \undefined \def \showISSN      #1{\unskip}     \fi
\ifx \showLCCN     \undefined \def \showLCCN      #1{\unskip}     \fi
\ifx \shownote     \undefined \def \shownote      #1{#1}          \fi
\ifx \showarticletitle \undefined \def \showarticletitle #1{#1}   \fi
\ifx \showURL      \undefined \def \showURL       {\relax}        \fi
% invisible to TeX
\providecommand\bibfield[2]{#2}
\providecommand\bibinfo[2]{#2}
\providecommand\natexlab[1]{#1}
\providecommand\showeprint[2][]{arXiv:#2}

\bibitem[\protect\citeauthoryear{Alshahwan, Gao, Harman, Jia, Mao, Mols, Tei,
  and Zorin}{Alshahwan et~al\mbox{.}}{2018}]%
        {open2}
\bibfield{author}{\bibinfo{person}{Nadia Alshahwan}, \bibinfo{person}{Xinbo
  Gao}, \bibinfo{person}{Mark Harman}, \bibinfo{person}{Yue Jia},
  \bibinfo{person}{Ke Mao}, \bibinfo{person}{Alexander Mols},
  \bibinfo{person}{Taijin Tei}, {and} \bibinfo{person}{Ilya Zorin}.}
  \bibinfo{year}{2018}\natexlab{}.
\newblock \showarticletitle{Deploying Search Based Software Engineering with
  Sapienz at Facebook}. In \bibinfo{booktitle}{\emph{Search-Based Software
  Engineering}}. \bibinfo{pages}{3--45}.
\newblock


\bibitem[\protect\citeauthoryear{Arora and Barak}{Arora and Barak}{2009}]%
        {sharpp}
\bibfield{author}{\bibinfo{person}{Sanjeev Arora} {and} \bibinfo{person}{Boaz
  Barak}.} \bibinfo{year}{2009}\natexlab{}.
\newblock \bibinfo{booktitle}{\emph{Computational Complexity: A Modern
  Approach} (\bibinfo{edition}{1st} ed.)}.
\newblock \bibinfo{publisher}{Cambridge University Press},
  \bibinfo{address}{New York, NY, USA}.
\newblock
\showISBNx{0521424267, 9780521424264}


\bibitem[\protect\citeauthoryear{Bang, Aydin, Phan, P\u{a}s\u{a}reanu, and
  Bultan}{Bang et~al\mbox{.}}{2016}]%
        {bultan}
\bibfield{author}{\bibinfo{person}{Lucas Bang}, \bibinfo{person}{Abdulbaki
  Aydin}, \bibinfo{person}{Quoc-Sang Phan}, \bibinfo{person}{Corina~S.
  P\u{a}s\u{a}reanu}, {and} \bibinfo{person}{Tevfik Bultan}.}
  \bibinfo{year}{2016}\natexlab{}.
\newblock \showarticletitle{String Analysis for Side Channels with Segmented
  Oracles}. In \bibinfo{booktitle}{\emph{Proceedings of the 2016 24th ACM
  SIGSOFT International Symposium on Foundations of Software Engineering}}
  \emph{(\bibinfo{series}{FSE 2016})}. \bibinfo{pages}{193--204}.
\newblock


\bibitem[\protect\citeauthoryear{B\"{o}hme}{B\"{o}hme}{2019}]%
        {assurances}
\bibfield{author}{\bibinfo{person}{Marcel B\"{o}hme}.}
  \bibinfo{year}{2019}\natexlab{}.
\newblock \showarticletitle{Assurance in Software Testing: A Roadmap}. In
  \bibinfo{booktitle}{\emph{Proceedings of the 41st International Conference on
  Software Engineering: New Ideas and Emerging Results}}
  \emph{(\bibinfo{series}{ICSE-NIER '19})}. \bibinfo{publisher}{IEEE Press},
  \bibinfo{address}{Piscataway, NJ, USA}, \bibinfo{pages}{5--8}.
\newblock


\bibitem[\protect\citeauthoryear{Borges, Filieri, D'Amorim, and
  P\u{a}s\u{a}reanu}{Borges et~al\mbox{.}}{2015}]%
        {filieri3}
\bibfield{author}{\bibinfo{person}{Mateus Borges}, \bibinfo{person}{Antonio
  Filieri}, \bibinfo{person}{Marcelo D'Amorim}, {and}
  \bibinfo{person}{Corina~S. P\u{a}s\u{a}reanu}.}
  \bibinfo{year}{2015}\natexlab{}.
\newblock \showarticletitle{Iterative Distribution-aware Sampling for
  Probabilistic Symbolic Execution}. In \bibinfo{booktitle}{\emph{Proceedings
  of the 2015 10th Joint Meeting on Foundations of Software Engineering}}
  \emph{(\bibinfo{series}{ESEC/FSE 2015})}. \bibinfo{pages}{866--877}.
\newblock


\bibitem[\protect\citeauthoryear{Borges, Filieri, D'Amorim, P\u{a}s\u{a}reanu,
  and Visser}{Borges et~al\mbox{.}}{2014}]%
        {filieri2}
\bibfield{author}{\bibinfo{person}{Mateus Borges}, \bibinfo{person}{Antonio
  Filieri}, \bibinfo{person}{Marcelo D'Amorim}, \bibinfo{person}{Corina~S.
  P\u{a}s\u{a}reanu}, {and} \bibinfo{person}{Willem Visser}.}
  \bibinfo{year}{2014}\natexlab{}.
\newblock \showarticletitle{Compositional Solution Space Quantification for
  Probabilistic Software Analysis}. In \bibinfo{booktitle}{\emph{Proceedings of
  the 35th ACM SIGPLAN Conference on Programming Language Design and
  Implementation}} \emph{(\bibinfo{series}{PLDI '14})}.
  \bibinfo{pages}{123--132}.
\newblock


\bibitem[\protect\citeauthoryear{Brennan, Tsiskaridze, Rosner, Aydin, and
  Bultan}{Brennan et~al\mbox{.}}{2017}]%
        {bultan2}
\bibfield{author}{\bibinfo{person}{Tegan Brennan}, \bibinfo{person}{Nestan
  Tsiskaridze}, \bibinfo{person}{Nicol\'{a}s Rosner},
  \bibinfo{person}{Abdulbaki Aydin}, {and} \bibinfo{person}{Tevfik Bultan}.}
  \bibinfo{year}{2017}\natexlab{}.
\newblock \showarticletitle{Constraint Normalization and Parameterized Caching
  for Quantitative Program Analysis}. In \bibinfo{booktitle}{\emph{Proceedings
  of the 2017 11th Joint Meeting on Foundations of Software Engineering}}
  \emph{(\bibinfo{series}{ESEC/FSE 2017})}. \bibinfo{pages}{535--546}.
\newblock


\bibitem[\protect\citeauthoryear{Brown, Cai, and DasGupta}{Brown
  et~al\mbox{.}}{2001}]%
        {onlyapprox}
\bibfield{author}{\bibinfo{person}{Lawrence~D. Brown}, \bibinfo{person}{T.~Tony
  Cai}, {and} \bibinfo{person}{Anirban DasGupta}.}
  \bibinfo{year}{2001}\natexlab{}.
\newblock \showarticletitle{Interval Estimation for a Binomial Proportion}.
\newblock \bibinfo{journal}{\emph{Statist. Sci.}} \bibinfo{volume}{16},
  \bibinfo{number}{2} (\bibinfo{date}{05} \bibinfo{year}{2001}),
  \bibinfo{pages}{101--133}.
\newblock


\bibitem[\protect\citeauthoryear{Calcagno, Distefano, O'Hearn, and
  Yang}{Calcagno et~al\mbox{.}}{2011}]%
        {infer}
\bibfield{author}{\bibinfo{person}{Cristiano Calcagno}, \bibinfo{person}{Dino
  Distefano}, \bibinfo{person}{Peter~W. O'Hearn}, {and}
  \bibinfo{person}{Hongseok Yang}.} \bibinfo{year}{2011}\natexlab{}.
\newblock \showarticletitle{Compositional Shape Analysis by Means of
  Bi-Abduction}.
\newblock \bibinfo{journal}{\emph{J. ACM}} \bibinfo{volume}{58},
  \bibinfo{number}{6}, Article \bibinfo{articleno}{26} (\bibinfo{date}{Dec.}
  \bibinfo{year}{2011}), \bibinfo{numpages}{66}~pages.
\newblock


\bibitem[\protect\citeauthoryear{Chen, Liu, and Le}{Chen et~al\mbox{.}}{2016}]%
        {bihuan}
\bibfield{author}{\bibinfo{person}{Bihuan Chen}, \bibinfo{person}{Yang Liu},
  {and} \bibinfo{person}{Wei Le}.} \bibinfo{year}{2016}\natexlab{}.
\newblock \showarticletitle{Generating Performance Distributions via
  Probabilistic Symbolic Execution}. In \bibinfo{booktitle}{\emph{Proceedings
  of the 38th International Conference on Software Engineering}}
  \emph{(\bibinfo{series}{ICSE '16})}. \bibinfo{pages}{49--60}.
\newblock


\bibitem[\protect\citeauthoryear{Chistikov, Dimitrova, and Majumdar}{Chistikov
  et~al\mbox{.}}{2017}]%
        {sharpSMT}
\bibfield{author}{\bibinfo{person}{Dmitry Chistikov}, \bibinfo{person}{Rayna
  Dimitrova}, {and} \bibinfo{person}{Rupak Majumdar}.}
  \bibinfo{year}{2017}\natexlab{}.
\newblock \showarticletitle{Approximate counting in SMT and value estimation
  for probabilistic programs}.
\newblock \bibinfo{journal}{\emph{Acta Informatica}} \bibinfo{volume}{54},
  \bibinfo{number}{8} (\bibinfo{date}{01 Dec} \bibinfo{year}{2017}),
  \bibinfo{pages}{729--764}.
\newblock


\bibitem[\protect\citeauthoryear{Clopper and Pearson}{Clopper and
  Pearson}{1934}]%
        {clopper}
\bibfield{author}{\bibinfo{person}{C.~J. Clopper} {and} \bibinfo{person}{E.~S.
  Pearson}.} \bibinfo{year}{1934}\natexlab{}.
\newblock \showarticletitle{The Use of Confidence or Fiducial Limits Illstrated
  in the Case of the Binomial}.
\newblock \bibinfo{journal}{\emph{Biometrika}} \bibinfo{volume}{26},
  \bibinfo{number}{4} (\bibinfo{date}{12} \bibinfo{year}{1934}),
  \bibinfo{pages}{404--413}.
\newblock


\bibitem[\protect\citeauthoryear{Dorai-Raj}{Dorai-Raj}{2015}]%
        {binom}
\bibfield{author}{\bibinfo{person}{Sundar Dorai-Raj}.}
  \bibinfo{year}{2015}\natexlab{}.
\newblock \bibinfo{title}{R -- Binom package}.
\newblock
  \bibinfo{howpublished}{\url{https://cran.r-project.org/web/packages/binom/binom.pdf}}.
\newblock
\newblock
\shownote{Accessed: 2019-06-17.}


\bibitem[\protect\citeauthoryear{Dum, Zoller, and Ritsch}{Dum
  et~al\mbox{.}}{1992}]%
        {mcappl}
\bibfield{author}{\bibinfo{person}{R. Dum}, \bibinfo{person}{P. Zoller}, {and}
  \bibinfo{person}{H. Ritsch}.} \bibinfo{year}{1992}\natexlab{}.
\newblock \showarticletitle{Monte Carlo simulation of the atomic master
  equation for spontaneous emission}.
\newblock \bibinfo{journal}{\emph{Physical Review A}}  \bibinfo{volume}{45}
  (\bibinfo{date}{Apr} \bibinfo{year}{1992}), \bibinfo{pages}{4879--4887}.
\newblock
Issue 7.


\bibitem[\protect\citeauthoryear{Filieri, Pasareanu, and Yang}{Filieri
  et~al\mbox{.}}{2015}]%
        {changes}
\bibfield{author}{\bibinfo{person}{A. Filieri}, \bibinfo{person}{C.~S.
  Pasareanu}, {and} \bibinfo{person}{G. Yang}.}
  \bibinfo{year}{2015}\natexlab{}.
\newblock \showarticletitle{Quantification of Software Changes through
  Probabilistic Symbolic Execution (N)}. In
  \bibinfo{booktitle}{\emph{Proceedings of the 30th IEEE/ACM International
  Conference on Automated Software Engineering}}
  \emph{(\bibinfo{series}{ASE'15})}. \bibinfo{pages}{703--708}.
\newblock


\bibitem[\protect\citeauthoryear{Filieri, P\u{a}s\u{a}reanu, and
  Visser}{Filieri et~al\mbox{.}}{2013}]%
        {filieri1}
\bibfield{author}{\bibinfo{person}{Antonio Filieri}, \bibinfo{person}{Corina~S.
  P\u{a}s\u{a}reanu}, {and} \bibinfo{person}{Willem Visser}.}
  \bibinfo{year}{2013}\natexlab{}.
\newblock \showarticletitle{Reliability Analysis in Symbolic Pathfinder}. In
  \bibinfo{booktitle}{\emph{Proceedings of the 2013 International Conference on
  Software Engineering}} \emph{(\bibinfo{series}{ICSE '13})}.
  \bibinfo{pages}{622--631}.
\newblock


\bibitem[\protect\citeauthoryear{Filieri, P\u{a}s\u{a}reanu, Visser, and
  Geldenhuys}{Filieri et~al\mbox{.}}{2014}]%
        {filieri4}
\bibfield{author}{\bibinfo{person}{Antonio Filieri}, \bibinfo{person}{Corina~S.
  P\u{a}s\u{a}reanu}, \bibinfo{person}{Willem Visser}, {and}
  \bibinfo{person}{Jaco Geldenhuys}.} \bibinfo{year}{2014}\natexlab{}.
\newblock \showarticletitle{Statistical Symbolic Execution with Informed
  Sampling}. In \bibinfo{booktitle}{\emph{Proceedings of the 22Nd ACM SIGSOFT
  International Symposium on Foundations of Software Engineering}}
  \emph{(\bibinfo{series}{FSE 2014})}. \bibinfo{pages}{437--448}.
\newblock


\bibitem[\protect\citeauthoryear{Geldenhuys, Dwyer, and Visser}{Geldenhuys
  et~al\mbox{.}}{2012}]%
        {pse}
\bibfield{author}{\bibinfo{person}{Jaco Geldenhuys},
  \bibinfo{person}{Matthew~B. Dwyer}, {and} \bibinfo{person}{Willem Visser}.}
  \bibinfo{year}{2012}\natexlab{}.
\newblock \showarticletitle{Probabilistic Symbolic Execution}. In
  \bibinfo{booktitle}{\emph{Proceedings of the 2012 International Symposium on
  Software Testing and Analysis}} \emph{(\bibinfo{series}{ISSTA 2012})}.
  \bibinfo{pages}{166--176}.
\newblock


\bibitem[\protect\citeauthoryear{Hammond, Lester, and Reynolds}{Hammond
  et~al\mbox{.}}{1994}]%
        {abinitio}
\bibfield{author}{\bibinfo{person}{Brian~L Hammond}, \bibinfo{person}{William~A
  Lester}, {and} \bibinfo{person}{Peter~James Reynolds}.}
  \bibinfo{year}{1994}\natexlab{}.
\newblock \bibinfo{booktitle}{\emph{Monte Carlo methods in ab initio quantum
  chemistry}}. Vol.~\bibinfo{volume}{1}.
\newblock \bibinfo{publisher}{World Scientific}.
\newblock


\bibitem[\protect\citeauthoryear{Hanley and Lippman-Hand}{Hanley and
  Lippman-Hand}{1983}]%
        {hanley}
\bibfield{author}{\bibinfo{person}{JA Hanley} {and} \bibinfo{person}{A
  Lippman-Hand}.} \bibinfo{year}{1983}\natexlab{}.
\newblock \showarticletitle{If nothing goes wrong, is everything all right?
  Interpreting zero numerators}.
\newblock \bibinfo{journal}{\emph{Journal of the American Medical Association}}
  \bibinfo{volume}{249}, \bibinfo{number}{13} (\bibinfo{year}{1983}),
  \bibinfo{pages}{1743--1745}.
\newblock


\bibitem[\protect\citeauthoryear{Harman and O'Hearn}{Harman and
  O'Hearn}{2018}]%
        {open1}
\bibfield{author}{\bibinfo{person}{Mark Harman} {and} \bibinfo{person}{Peter
  O'Hearn}.} \bibinfo{year}{2018}\natexlab{}.
\newblock \bibinfo{title}{From Start-ups to Scale-ups: Opportunities and Open
  Problems for Static and Dynamic Program Analysis}.
\newblock
\newblock
\newblock
\shownote{Keynote at the 18th IEEE International Working Conference on Source
  Code Analysis.}


\bibitem[\protect\citeauthoryear{H\'erault, Lassaigne, Magniette, and
  Peyronnet}{H\'erault et~al\mbox{.}}{2004}]%
        {probmodelcheck1}
\bibfield{author}{\bibinfo{person}{T. H\'erault}, \bibinfo{person}{R.
  Lassaigne}, \bibinfo{person}{F. Magniette}, {and} \bibinfo{person}{S.
  Peyronnet}.} \bibinfo{year}{2004}\natexlab{}.
\newblock \showarticletitle{Approximate Probabilistic Model Checking}. In
  \bibinfo{booktitle}{\emph{Proceedings of the 5th International Conference on
  Verification, Model Checking and Abstract Interpretation}}
  \emph{(\bibinfo{series}{VMCAI'04})}. \bibinfo{pages}{307--329}.
\newblock


\bibitem[\protect\citeauthoryear{Hoeffding}{Hoeffding}{1963}]%
        {hoeff}
\bibfield{author}{\bibinfo{person}{Wassily Hoeffding}.}
  \bibinfo{year}{1963}\natexlab{}.
\newblock \showarticletitle{Probability Inequalities for Sums of Bounded Random
  Variables}.
\newblock \bibinfo{journal}{\emph{J. Amer. Statist. Assoc.}}
  \bibinfo{volume}{58}, \bibinfo{number}{301} (\bibinfo{year}{1963}),
  \bibinfo{pages}{13--30}.
\newblock


\bibitem[\protect\citeauthoryear{Katoen}{Katoen}{2016}]%
        {probmodelcheck2}
\bibfield{author}{\bibinfo{person}{Joost-Pieter Katoen}.}
  \bibinfo{year}{2016}\natexlab{}.
\newblock \showarticletitle{The Probabilistic Model Checking Landscape}. In
  \bibinfo{booktitle}{\emph{Proceedings of the 31st Annual ACM/IEEE Symposium
  on Logic in Computer Science}} \emph{(\bibinfo{series}{LICS '16})}.
  \bibinfo{pages}{31--45}.
\newblock


\bibitem[\protect\citeauthoryear{Kearns and Vazirani}{Kearns and
  Vazirani}{1994}]%
        {paclearning}
\bibfield{author}{\bibinfo{person}{Michael~J. Kearns} {and}
  \bibinfo{person}{Umesh~V. Vazirani}.} \bibinfo{year}{1994}\natexlab{}.
\newblock \bibinfo{booktitle}{\emph{An Introduction to Computational Learning
  Theory}}.
\newblock \bibinfo{publisher}{MIT Press}, \bibinfo{address}{Cambridge, MA,
  USA}.
\newblock
\showISBNx{0-262-11193-4}


\bibitem[\protect\citeauthoryear{Kohavi and Longbotham}{Kohavi and
  Longbotham}{2017}]%
        {abtesting}
\bibfield{author}{\bibinfo{person}{Ron Kohavi} {and} \bibinfo{person}{Roger
  Longbotham}.} \bibinfo{year}{2017}\natexlab{}.
\newblock \bibinfo{booktitle}{\emph{Online Controlled Experiments and A/B
  Testing}}.
\newblock \bibinfo{publisher}{Springer US}, \bibinfo{pages}{922--929}.
\newblock


\bibitem[\protect\citeauthoryear{Krishnamoorthy and Peng}{Krishnamoorthy and
  Peng}{2007}]%
        {predictclopper}
\bibfield{author}{\bibinfo{person}{K. Krishnamoorthy} {and}
  \bibinfo{person}{Jie Peng}.} \bibinfo{year}{2007}\natexlab{}.
\newblock \showarticletitle{Some Properties of the Exact and Score Methods for
  Binomial Proportion and Sample Size Calculation}.
\newblock \bibinfo{journal}{\emph{Communications in Statistics - Simulation and
  Computation}} \bibinfo{volume}{36}, \bibinfo{number}{6}
  (\bibinfo{year}{2007}), \bibinfo{pages}{1171--1186}.
\newblock


\bibitem[\protect\citeauthoryear{Kwiatkowska, Norman, and Parker}{Kwiatkowska
  et~al\mbox{.}}{2011}]%
        {pmc}
\bibfield{author}{\bibinfo{person}{Marta Kwiatkowska}, \bibinfo{person}{Gethin
  Norman}, {and} \bibinfo{person}{David Parker}.}
  \bibinfo{year}{2011}\natexlab{}.
\newblock \showarticletitle{PRISM 4.0: Verification of Probabilistic Real-time
  Systems}. In \bibinfo{booktitle}{\emph{Proceedings of the 23rd International
  Conference on Computer Aided Verification}}
  \emph{(\bibinfo{series}{CAV'11})}. \bibinfo{pages}{585--591}.
\newblock


\bibitem[\protect\citeauthoryear{Legay, Delahaye, and Bensalem}{Legay
  et~al\mbox{.}}{2010}]%
        {smc}
\bibfield{author}{\bibinfo{person}{Axel Legay}, \bibinfo{person}{Beno\^{\i}t
  Delahaye}, {and} \bibinfo{person}{Saddek Bensalem}.}
  \bibinfo{year}{2010}\natexlab{}.
\newblock \showarticletitle{Statistical Model Checking: An Overview}. In
  \bibinfo{booktitle}{\emph{Proceedings of the First International Conference
  on Runtime Verification}} \emph{(\bibinfo{series}{RV'10})}.
  \bibinfo{pages}{122--135}.
\newblock


\bibitem[\protect\citeauthoryear{Llerena, B\"{o}hme, Br\"{u}nink, Su, and
  Rosenblum}{Llerena et~al\mbox{.}}{2018}]%
        {yamilet}
\bibfield{author}{\bibinfo{person}{Yamilet R.~Serrano Llerena},
  \bibinfo{person}{Marcel B\"{o}hme}, \bibinfo{person}{Marc Br\"{u}nink},
  \bibinfo{person}{Guoxin Su}, {and} \bibinfo{person}{David~S. Rosenblum}.}
  \bibinfo{year}{2018}\natexlab{}.
\newblock \showarticletitle{Verifying the Long-Run Behavior of Probabilistic
  System Models in the Presence of Uncertainty}. In
  \bibinfo{booktitle}{\emph{Proceedings of the 12th Joint meeting of the
  European Software Engineering Conference and the ACM SIGSOFT Symposium on the
  Foundations of Software Engineering}} \emph{(\bibinfo{series}{ESEC/FSE})}.
  \bibinfo{pages}{1--11}.
\newblock


\bibitem[\protect\citeauthoryear{Loera, Hemmecke, Tauzer, and Yoshida}{Loera
  et~al\mbox{.}}{2004}]%
        {latte}
\bibfield{author}{\bibinfo{person}{Jes\'us A.~De Loera},
  \bibinfo{person}{Raymond Hemmecke}, \bibinfo{person}{Jeremiah Tauzer}, {and}
  \bibinfo{person}{Ruriko Yoshida}.} \bibinfo{year}{2004}\natexlab{}.
\newblock \showarticletitle{Effective lattice point counting in rational convex
  polytopes}.
\newblock \bibinfo{journal}{\emph{Journal of Symbolic Computation}}
  \bibinfo{volume}{38}, \bibinfo{number}{4} (\bibinfo{year}{2004}),
  \bibinfo{pages}{1273 -- 1302}.
\newblock


\bibitem[\protect\citeauthoryear{Luckow, P\u{a}s\u{a}reanu, Dwyer, Filieri, and
  Visser}{Luckow et~al\mbox{.}}{2014}]%
        {filieri5}
\bibfield{author}{\bibinfo{person}{Kasper Luckow}, \bibinfo{person}{Corina~S.
  P\u{a}s\u{a}reanu}, \bibinfo{person}{Matthew~B. Dwyer},
  \bibinfo{person}{Antonio Filieri}, {and} \bibinfo{person}{Willem Visser}.}
  \bibinfo{year}{2014}\natexlab{}.
\newblock \showarticletitle{Exact and Approximate Probabilistic Symbolic
  Execution for Nondeterministic Programs}. In
  \bibinfo{booktitle}{\emph{Proceedings of the 29th ACM/IEEE International
  Conference on Automated Software Engineering}} \emph{(\bibinfo{series}{ASE
  '14})}. \bibinfo{pages}{575--586}.
\newblock


\bibitem[\protect\citeauthoryear{M{\o}lmer, Castin, and Dalibard}{M{\o}lmer
  et~al\mbox{.}}{1993}]%
        {cannot}
\bibfield{author}{\bibinfo{person}{Klaus M{\o}lmer}, \bibinfo{person}{Yvan
  Castin}, {and} \bibinfo{person}{Jean Dalibard}.}
  \bibinfo{year}{1993}\natexlab{}.
\newblock \showarticletitle{Monte Carlo wave-function method in quantum
  optics}.
\newblock \bibinfo{journal}{\emph{Journal of the Optical Society of America B}}
  \bibinfo{volume}{10}, \bibinfo{number}{3} (\bibinfo{date}{Mar}
  \bibinfo{year}{1993}), \bibinfo{pages}{524--538}.
\newblock


\bibitem[\protect\citeauthoryear{Phan, Malacaria, P\u{a}s\u{a}reanu, and
  D'Amorim}{Phan et~al\mbox{.}}{2014}]%
        {qif}
\bibfield{author}{\bibinfo{person}{Quoc-Sang Phan}, \bibinfo{person}{Pasquale
  Malacaria}, \bibinfo{person}{Corina~S. P\u{a}s\u{a}reanu}, {and}
  \bibinfo{person}{Marcelo D'Amorim}.} \bibinfo{year}{2014}\natexlab{}.
\newblock \showarticletitle{Quantifying Information Leaks Using Reliability
  Analysis}. In \bibinfo{booktitle}{\emph{Proceedings of the 2014 International
  SPIN Symposium on Model Checking of Software}} \emph{(\bibinfo{series}{SPIN
  2014})}. \bibinfo{pages}{105--108}.
\newblock


\bibitem[\protect\citeauthoryear{Ramalingam}{Ramalingam}{1994}]%
        {alias}
\bibfield{author}{\bibinfo{person}{G. Ramalingam}.}
  \bibinfo{year}{1994}\natexlab{}.
\newblock \showarticletitle{The Undecidability of Aliasing}.
\newblock \bibinfo{journal}{\emph{ACM Trans. Program. Lang. Syst.}}
  \bibinfo{volume}{16}, \bibinfo{number}{5} (\bibinfo{date}{Sept.}
  \bibinfo{year}{1994}), \bibinfo{pages}{1467--1471}.
\newblock


\bibitem[\protect\citeauthoryear{{Rosenblum}}{{Rosenblum}}{2016}]%
        {probThinking}
\bibfield{author}{\bibinfo{person}{D.~S. {Rosenblum}}.}
  \bibinfo{year}{2016}\natexlab{}.
\newblock \showarticletitle{The power of probabilistic thinking}. In
  \bibinfo{booktitle}{\emph{2016 31st IEEE/ACM International Conference on
  Automated Software Engineering (ASE)}}. \bibinfo{pages}{3--3}.
\newblock


\bibitem[\protect\citeauthoryear{Sadowski, Aftandilian, Eagle, Miller-Cushon,
  and Jaspan}{Sadowski et~al\mbox{.}}{2018}]%
        {errorprone}
\bibfield{author}{\bibinfo{person}{Caitlin Sadowski}, \bibinfo{person}{Edward
  Aftandilian}, \bibinfo{person}{Alex Eagle}, \bibinfo{person}{Liam
  Miller-Cushon}, {and} \bibinfo{person}{Ciera Jaspan}.}
  \bibinfo{year}{2018}\natexlab{}.
\newblock \showarticletitle{Lessons from Building Static Analysis Tools at
  Google}.
\newblock \bibinfo{journal}{\emph{Commun. ACM}} \bibinfo{volume}{61},
  \bibinfo{number}{4} (\bibinfo{date}{March} \bibinfo{year}{2018}),
  \bibinfo{pages}{58--66}.
\newblock


\bibitem[\protect\citeauthoryear{Sankaranarayanan, Chakarov, and
  Gulwani}{Sankaranarayanan et~al\mbox{.}}{2013}]%
        {gulwani}
\bibfield{author}{\bibinfo{person}{Sriram Sankaranarayanan},
  \bibinfo{person}{Aleksandar Chakarov}, {and} \bibinfo{person}{Sumit
  Gulwani}.} \bibinfo{year}{2013}\natexlab{}.
\newblock \showarticletitle{Static Analysis for Probabilistic Programs:
  Inferring Whole Program Properties from Finitely Many Paths}. In
  \bibinfo{booktitle}{\emph{Proceedings of the 34th ACM SIGPLAN Conference on
  Programming Language Design and Implementation}} \emph{(\bibinfo{series}{PLDI
  '13})}. \bibinfo{pages}{447--458}.
\newblock


\bibitem[\protect\citeauthoryear{Sen, Viswanathan, and Agha}{Sen
  et~al\mbox{.}}{2004}]%
        {sen}
\bibfield{author}{\bibinfo{person}{Koushik Sen}, \bibinfo{person}{Mahesh
  Viswanathan}, {and} \bibinfo{person}{Gul Agha}.}
  \bibinfo{year}{2004}\natexlab{}.
\newblock \showarticletitle{Statistical Model Checking of Black-Box
  Probabilistic Systems}. In \bibinfo{booktitle}{\emph{Computer Aided
  Verification}}. \bibinfo{pages}{202--215}.
\newblock


\bibitem[\protect\citeauthoryear{Serebryany, Bruening, Potapenko, and
  Vyukov}{Serebryany et~al\mbox{.}}{2012}]%
        {asan}
\bibfield{author}{\bibinfo{person}{Konstantin Serebryany},
  \bibinfo{person}{Derek Bruening}, \bibinfo{person}{Alexander Potapenko},
  {and} \bibinfo{person}{Dmitry Vyukov}.} \bibinfo{year}{2012}\natexlab{}.
\newblock \showarticletitle{AddressSanitizer: A Fast Address Sanity Checker}.
  In \bibinfo{booktitle}{\emph{Proceedings of the 2012 USENIX Conference on
  Annual Technical Conference}} \emph{(\bibinfo{series}{USENIX ATC'12})}.
  \bibinfo{publisher}{USENIX Association}, \bibinfo{address}{Berkeley, CA,
  USA}, \bibinfo{pages}{28--28}.
\newblock


\bibitem[\protect\citeauthoryear{Shalev-Shwartz and Ben-David}{Shalev-Shwartz
  and Ben-David}{2014}]%
        {mlshalev}
\bibfield{author}{\bibinfo{person}{Shai Shalev-Shwartz} {and}
  \bibinfo{person}{Shai Ben-David}.} \bibinfo{year}{2014}\natexlab{}.
\newblock \bibinfo{booktitle}{\emph{Understanding Machine Learning: From Theory
  to Algorithms}}.
\newblock \bibinfo{publisher}{Cambridge University Press},
  \bibinfo{address}{New York, NY, USA}.
\newblock
\showISBNx{1107057132, 9781107057135}


\bibitem[\protect\citeauthoryear{Valiant}{Valiant}{2013}]%
        {valiant}
\bibfield{author}{\bibinfo{person}{Leslie Valiant}.}
  \bibinfo{year}{2013}\natexlab{}.
\newblock \bibinfo{booktitle}{\emph{Probably Approximately Correct: Nature's
  Algorithms for Learning and Prospering in a Complex World}}.
\newblock \bibinfo{publisher}{Basic Books, Inc.}
\newblock
\showISBNx{0465032710, 9780465032716}


\bibitem[\protect\citeauthoryear{Valiant}{Valiant}{1984}]%
        {valiant84}
\bibfield{author}{\bibinfo{person}{L.~G. Valiant}.}
  \bibinfo{year}{1984}\natexlab{}.
\newblock \showarticletitle{A Theory of the Learnable}.
\newblock \bibinfo{journal}{\emph{Commun. ACM}} \bibinfo{volume}{27},
  \bibinfo{number}{11} (\bibinfo{date}{Nov.} \bibinfo{year}{1984}),
  \bibinfo{pages}{1134--1142}.
\newblock


\bibitem[\protect\citeauthoryear{Visser}{Visser}{2017}]%
        {lpar}
\bibfield{author}{\bibinfo{person}{Willem Visser}.}
  \bibinfo{year}{2017}\natexlab{}.
\newblock \bibinfo{title}{Probabilistic Symbolic Execution: A New Hammer}.
\newblock
\newblock
\urldef\tempurl%
\url{https://easychair.org/smart-program/LPAR-21/LPAR-21-s3-Visser.pdf}
\showURL{%
\tempurl}
\newblock
\shownote{Keynote at the 21st International Conference on Logic for
  Programming, Artificial Intelligence and Reasoning (LPAR-21).}


\bibitem[\protect\citeauthoryear{Website}{Website}{2010}]%
        {visa}
\bibfield{author}{\bibinfo{person}{Website}.} \bibinfo{year}{2010}\natexlab{}.
\newblock \bibinfo{title}{{Visa: Transactions per Day}}.
\newblock
  \bibinfo{howpublished}{\url{https://usa.visa.com/run-your-business/small-business-tools/retail.html}}.
\newblock
\newblock
\shownote{Accessed: 2019-06-17.}


\bibitem[\protect\citeauthoryear{Website}{Website}{2017}]%
        {oss}
\bibfield{author}{\bibinfo{person}{Website}.} \bibinfo{year}{2017}\natexlab{}.
\newblock \bibinfo{title}{{OSS-Fuzz: Five Months Later}}.
\newblock
  \bibinfo{howpublished}{\url{https://testing.googleblog.com/2017/05/oss-fuzz-five-months-later-and.html}}.
\newblock
\newblock
\shownote{Accessed: 2017-11-13.}


\bibitem[\protect\citeauthoryear{Website}{Website}{2018a}]%
        {livestats2}
\bibfield{author}{\bibinfo{person}{Website}.} \bibinfo{year}{2018}\natexlab{a}.
\newblock \bibinfo{title}{Domo: Data Never Sleeps Report 6.0}.
\newblock
  \bibinfo{howpublished}{\url{https://www.domo.com/blog/data-never-sleeps-6/}}.
\newblock
\newblock
\shownote{Accessed: 2018-11-16.}


\bibitem[\protect\citeauthoryear{Website}{Website}{2018b}]%
        {fbStat}
\bibfield{author}{\bibinfo{person}{Website}.} \bibinfo{year}{2018}\natexlab{b}.
\newblock \bibinfo{title}{Facebook: Company Info and Statistics}.
\newblock \bibinfo{howpublished}{\url{https://newsroom.fb.com/company-info/}}.
\newblock
\newblock
\shownote{Accessed: 2018-11-16.}


\bibitem[\protect\citeauthoryear{Website}{Website}{2018c}]%
        {netflix}
\bibfield{author}{\bibinfo{person}{Website}.} \bibinfo{year}{2018}\natexlab{c}.
\newblock \bibinfo{title}{Netflix Tech Blog: Automated Canary Analysis at
  Netflix with Kayenta}.
\newblock
  \bibinfo{howpublished}{\url{https://medium.com/netflix-techblog/automated-canary-analysis-at-netflix-with-kayenta-3260bc7acc69}}.
\newblock
\newblock
\shownote{Accessed: 2018-10-13.}


\bibitem[\protect\citeauthoryear{Website}{Website}{2018d}]%
        {netflixStat}
\bibfield{author}{\bibinfo{person}{Website}.} \bibinfo{year}{2018}\natexlab{d}.
\newblock \bibinfo{title}{Netflix TechBlog: Edge Load Balancing}.
\newblock
  \bibinfo{howpublished}{\url{https://medium.com/netflix-techblog/netflix-edge-load-balancing-695308b5548c}}.
\newblock
\newblock
\shownote{Accessed: 2018-11-16.}


\bibitem[\protect\citeauthoryear{Website}{Website}{2019}]%
        {tinder}
\bibfield{author}{\bibinfo{person}{Website}.} \bibinfo{year}{2019}\natexlab{}.
\newblock \bibinfo{title}{{Visa: Swipes per Day}}.
\newblock
  \bibinfo{howpublished}{\url{https://www.gotinder.com/press?locale=en}}.
\newblock
\newblock
\shownote{Accessed: 2019-06-17.}


\bibitem[\protect\citeauthoryear{Wilson}{Wilson}{1927}]%
        {wilson}
\bibfield{author}{\bibinfo{person}{Edwin~B. Wilson}.}
  \bibinfo{year}{1927}\natexlab{}.
\newblock \showarticletitle{Probable Inference, the Law of Succession, and
  Statistical Inference}.
\newblock \bibinfo{journal}{\emph{J. Amer. Statist. Assoc.}}
  \bibinfo{volume}{22}, \bibinfo{number}{158} (\bibinfo{year}{1927}),
  \bibinfo{pages}{209--212}.
\newblock


\end{thebibliography}

%% Appendix
%\appendix
%\input{sections/appendix}
%\newpage
%\input{sections/categorical}
 
\end{document}